\documentclass[%
reprint, superscriptaddress,
 amsmath,amssymb,
 aps,
]{revtex4-1}

\usepackage[english]{babel}
\usepackage[utf8]{inputenc}

\setcitestyle{sort&compress,numbers}
\usepackage{amsthm, color, mathtools, amsmath, amssymb, setspace,bbm, tensor, subfigure, mathtools, placeins, graphicx, wrapfig,float, verbatim}
\usepackage{graphicx}
\usepackage[unicode=true, breaklinks=false, pdfborder={0 0 1}, backref=false, colorlinks=true, linkcolor=blue, citecolor=blue]{hyperref}

\usepackage{cancel,bm}

\let\oldsqrt\sqrt
\def\sqrt{\mathpalette\DHLhksqrt}
\def\DHLhksqrt#1#2{%
\setbox0=\hbox{$#1\oldsqrt{#2\,}$}\dimen0=\ht0
\advance\dimen0-0.2\ht0
\setbox2=\hbox{\vrule height\ht0 depth -\dimen0}%
{\box0\lower0.4pt\box2}}

\newcommand{\tr}{\operatorname{tr}}
\newcommand{\Tcal}{\mathcal{T}}

\def\bra#1{\mathinner{\langle{#1}|}}

\def\ket#1{\mathinner{|{#1}\rangle}}

\def\braket#1{\mathinner{\langle{#1}\rangle}}

\newcommand*\xbar[1]{%
   \hbox{%
     \vbox{%
       \hrule height 0.5pt % The actual bar
       \kern0.5ex%         % Distance between bar and symbol
       \hbox{%
         \kern-0.2em%      % Shortening on the left side
         \ensuremath{#1}%
         \kern-0.0em%      % Shortening on the right side
       }%
     }%
   }%
}

\def\BraVert{\egroup\,\mid\,\bgroup}

\def\ketbra#1#2{\ket{#1\vphantom{#2}}\!\bra{#2\vphantom{#1}}}
\def\bra#1{\mathinner{\langle{#1}|}}

\def\ket#1{\mathinner{|{#1}\rangle}}

\def\braket#1{\mathinner{\langle{#1}\rangle}}

\usepackage{bbm, braket}
\usepackage[mathscr]{euscript}
\allowdisplaybreaks
\newtheorem*{theorem*}{Theorem}

\begin{document}

\title{Reconstructing non-Markovian quantum dynamics with limited control}

\author{Simon Milz}
    \affiliation{School of Physics and Astronomy, Monash University, Clayton, Victoria 3800, Australia.}
    \email{simon.milz@monash.edu}

\author{Felix A. Pollock}
    \affiliation{School of Physics and Astronomy, Monash University, Clayton, Victoria 3800, Australia.}
    \email{felix.pollock@monash.edu}

\author{Kavan Modi}
    \affiliation{School of Physics and Astronomy, Monash University, Clayton, Victoria 3800, Australia.}
    \email{kavan.modi@monash.edu}

\date{\today}

\begin{abstract}
The dynamics of an open quantum system can be fully described and tomographically reconstructed if the experimenter has complete control over the system of interest. Most real-world experiments do not fulfill this assumption, and the amount of control is restricted by the respective experimental setup. That is, the set of performable manipulations of the system is limited. In this paper, we provide a general reconstruction scheme that yields an operationally well-defined description of non-Markovian quantum dynamics for a large class of realistic experimental situations. The resultant `restricted' descriptor for the process, surprisingly, does not have the property of complete positivity. Based on these restricted dynamics, we construct witnesses for correlations and the presence of memory effects. We demonstrate the applicability of our framework for the two important cases where the set of performable operations comprises only unitary operations or projective measurements, respectively, and show that it provides a powerful tool for the description of quantum control experiments.

\end{abstract}
%\pacs{asdfsadf}
\maketitle

\section{Introduction}
Traditionally, the dynamics of open quantum systems, \textit{i.e.}, systems that are coupled to an environment, are found by solving an equation of motion for the reduced state of the former, and are described by channels that map initial states of the system to final states~\footnote{Unless otherwise specified, when we refer to a system's state, we mean its density operator.}. Such methods only account for two-time correlation functions, which is sufficient to describe Markov processes~\cite{PARRAVICINI1977423,nielsen_quantum_2000, chuang_prescription_1997, breuer_theory_2007}, but generally insufficient to fully describe an arbitrary (non-Markovian) quantum process~\cite{breuer_theory_2007, modi_preparation_2011, milz_introduction_2017}.

To overcome this problem, and to provide a description of non-Markovian processes that can properly account for multi-time correlation functions, an operational framework, the so-called process tensor formalism, was recently introduced in~\cite{pollock_operational_prl, pollock_complete_pra, 1367-2630-18-6-063032}. This framework takes into account that an experimenter often has to measure -- or, more generally, manipulate -- the system of interest at several times during its evolution before determining its final state. The process tensor is the multilinear mapping from the operations the experimenter implemented on the system to the final state after the experiment has concluded.

The process tensor framework thereby remedies two shortcomings of traditional descriptions of open quantum system dynamics; it captures \textit{all} non-Markovian features (the memory) in any quantum process, and accounts for more general experimental scenarios. Indeed, this formalism provides unambiguous sufficient \textit{and} necessary criteria to test if a quantum process is non-Markovian. 
On a more fundamental level, the process tensor formalism naturally leads to a generalization of the Kolmogorov extension theorem~\cite{kolmogorov_foundations_1956, breuer_theory_2007} -- the theorem that defines the notion and ensures the existence of classical stochastic processes -- to the quantum case~\cite{milz_kolmogorov_2017}, putting quantum stochastic processes on the same foundational footing as their classical counterparts. Furthermore, it is a completely positive (CP) map, a property that helps to resolve paradoxical violations of fundamental physical and information-theoretic bounds~\cite{modi_role_2010, PhysRevLett.113.140502, argentieri2014violations, holevo-assingment} by constructing bounds that properly account for the initial correlations between the system and its environment~\cite{PhysRevA.92.052310}. 

In order to reconstruct the process tensor experimentally, \textit{i.e.}, in order to collect all accessible information about the underlying dynamics, it is necessary to be able to perform manipulations of the system that constitute a basis of all physical operations, that is, a basis of the set of (trace non-increasing) CP maps acting on the system of interest. This requirement of full local control is not met in most real-world experiments. The present work is an extension of this formalism to cases with limited experimental control.

Generally, experimenters are only interested in the effects of a restricted set of manipulations, or the set of available manipulations is fundamentally limited by the experimental setup. Well-known examples are ubiquitous in the field of quantum control~\cite{altafini_modeling_2012}, where, \textit{e.g.}, shaped laser pulses are used to control the dynamics of molecules~\cite{shapiro_principles_2003} or partial measurements are employed to steer the system of interest to a desired final state~\cite{blok_manipulating_2014}. Other prominent examples include dynamical decoupling experiments~\cite{viola_dynamical_1999}, and experimental setups in quantum optics, where beam splitters and phase shifters can be used to implement arbitrary (single-photon) unitary gates~\cite{reck_experimental_1994}, but no non-unitary operations.

A concrete and important example of such a scenario is the publicly available quantum computer by IBM~\cite{IBM}. This computer only allows for a sequence of one and two-qubit unitary gates followed by a final measurement on each qubit; \textit{a priori}, it is not obvious how to experimentally reconstruct a map that yields the correct output states for any sequence of unitary operations. Such a reconstruction could play an important role in building a model of (non-Markovian) errors, and measuring the output states corresponding to only a \emph{finite} set of such sequences would still yield some information about the dynamics of the computer. The resulting \textit{restricted} process tensor would then contain the multi-time correlations that can be probed by the available set of experimental operations.

All of the experimental procedures listed above can readily and meaningfully be cast in the language of the process tensor formalism. However, neither the influence of laser pulses, nor partial measurements, nor unitary operations, when taken as distinct sets, constitute a basis of the set of all possible manipulations of the system of interest. To make matters worse, these operations do not even constitute a convex space, let alone a linear vector space. It is, therefore, unclear if a meaningful process tensor can be experimentally reconstructed from this limited set of accessible manipulations, where by meaningful we mean that it maps every realizable input manipulation to the correct corresponding output state.

Moreover, we can ask what can be inferred about correlations and memory effects based on restricted process tensors. While the complete process tensor allows one to determine whether there are detectable correlations between the system and its environment, and/or memory effects in the dynamics~\cite{modi_operational_2012, pollock_complete_pra, pollock_operational_prl}, it is not obvious how to make similar assertions based on only a limited set of available experimental operations.

In this paper, we consider the most general scenario for restricted process tensors. We answer three basic questions: With only limited control, how can one reconstruct a meaningful description of multi-step experiments and non-Markovian dynamics? What properties does the reconstructed description have? And finally, can one make assertions about correlations and/or memory effects based on this reconstructed \textit{restricted} process tensor? 

Refs.~\cite{ziman_reconstruction_2004, ziman_process_2005} have investigated similar questions for the reconstruction of quantum channels based on incomplete information, while Ref.~\cite{kuah_how_2007} examined a question closely aligned with the present work. There, the authors considered the dynamics of a qubit, that is initially correlated with its environment, subsequent to a projective preparation. Their constructed restricted map -- a particular case of a restricted process tensor -- is well-defined, \textit{i.e.}, yields the correct output state, on any projective preparation of the qubit. 

Here, we begin by introducing process tensors and their experimental reconstruction in Sec.~\ref{subsec::ReCon}. We then modify this reconstruction procedure to account for the restrictions imposed by given experimental setups in Sec.~\ref{subsec::Restr}. The applicability of the introduced concepts is illustrated in detail in Sec.~\ref{subsec::UnitProj} for two extremal and experimentally relevant cases: where only unitary operations and only projective measurements are available to the experimenter, respectively. In Sec.~\ref{subsec::RestrProp} we analyze the properties and realm of validity of restricted process tensors. It turns out -- somewhat surprisingly -- that restricted process tensors can even describe situation that lie outside of what can be experimentally implemented. Unlike the full process tensor, they are however generally not completely positive. In Secs.~\ref{subsec::InitCorrWit} and~\ref{subsec::NonMark}, we construct witnesses for initial system-environment correlations and the non-Markovianity of processes based on restricted process tensors. Such witnesses can always be constructed, regardless of the limitations set by the experimental situation. Finally, in Sec.~\ref{sec::Examples} we supply concrete examples for the reconstruction of restricted process tensors and apply them to the field of quantum control of open systems~\cite{altafini_modeling_2012}. In particular, we investigate the case of dynamical decoupling~\cite{viola_dynamical_1999}, and show that restricted process tensors provide the right tool to capture the influence of local Hamiltonians on the dynamics of an open system when no microscopic model can be assumed.

Finally, while the process tensor formalism was developed for the description of general open quantum processes, it is closely related, mathematically, to the concepts of supermaps~\cite{chiribella_transforming_2008}, quantum combs~\cite{chiribella_quantum_2008}, causal modelling~\cite{1367-2630-18-6-063032, oreshkov_causal_2016, allen_quantum_2017}, causal boxes~\cite{portmann_causal_2015} and the operator tensor formulation of quantum theory~\cite{hardy_operator_2012, hardy_operational_2016}. The present work therefore applies to all of these settings when allowed control operations do not span the full space of those that are mathematically allowed.
 
\section{Background}
\label{subsec::ReCon}
The most general operationally meaningful description of any quantum process is a map from experimentally controllable inputs to final output states, which can be determined by means of quantum state tomography (QST)~\cite{nielsen_quantum_2000, chuang_prescription_1997, poyatos_complete_1997}. Depending on the experimental setup in question, the inputs could be initial system states, initial preparations, sequences of local operations, or both initial states \textit{and} sequences of local operations. Making use of the linearity of quantum mechanics, for each of these cases, the map describing the process can experimentally be reconstructed by measuring the final states corresponding to a complete basis of the inputs~\footnote{Note that far more general inputs and outputs can be considered~\cite{chiribella_transforming_2008,chiribella_quantum_2008}, but for the experimental setups we aim to describe, the aforementioned list is sufficient.}. We emphasize that such a map contains all relevant statistical data about the underlying process, \textit{e.g.}, outcome probabilities of sequential measurements. 

It is instructive, as a first step, to recall the well-known case of quantum channels $\hat \Lambda: \rho \mapsto \rho'$  where initial (final) system states are considered to be the inputs (outputs) of the process. This description is adequate if only two-time correlations, \textit{i.e.}, between the initial state -- prepared by the experimenter at $t_0$ -- and the final state -- measured by the experimenter at a later time $t_1$ -- are of interest. Due to linearity $\hat \Lambda$ is unambiguously defined by its action on a basis $\left\{\rho_\alpha\right\}_{\alpha=1}^{d_{\mathcal{S}}^2}$ of $\mathcal{B}(\mathcal{H}_\mathcal{S})$, where $\mathcal{B}(\mathcal{H}_\mathcal{S})$ is the set of bounded linear operators on the system Hilbert space $\mathcal{H}_\mathcal{S}$ and $d_{\mathcal{S}}$ is the dimension of $\mathcal{H}_\mathcal{S}$. Every input state $\rho$ can be decomposed as $\rho = \sum_{\alpha=1} r_\alpha \rho_\alpha$, and hence the action of $\hat \Lambda$ on $\rho$ is given by $\hat \Lambda[\rho] = \sum_\alpha r_\alpha \hat \Lambda[\rho_\alpha] := \sum_\alpha r_\alpha \rho_\alpha'$; once the output states $\rho_\alpha' = \hat \Lambda[\rho_\alpha]$ for a basis of input states are known, the map describing the underlying process is entirely defined. This fact forms the basis of process tomography (see, \textit{e.g.}, Ref.~\cite{chuang_prescription_1997}) where the map $\hat \Lambda$ is reconstructed experimentally by determining the output states $\rho_\alpha'=\Lambda\left[\rho_\alpha\right]$ for a set of $d_\mathcal{S}^2$ linearly independent input states and employing linear inversion techniques. 

While there are different equivalent ways to explicitly represent quantum maps (see, \textit{e.g.}, Ref.~\cite{milz_introduction_2017} for a detailed discussion of representations of quantum maps), in this paper we will solely employ the Choi representation~\cite{jamiolkowski_linear_1972,choi75} (also known as the B-form~\cite{jordan_dynamical_1961, SudarshanMatthewsRau61}). In this representation, every CP map $\hat \Lambda$ is mapped onto a positive matrix $\Lambda$, by letting $\hat \Lambda$ act on one half of an unnormalized maximally entangled state $\ket{\Phi^+} = \sum_{i=1}^{d_\mathcal{S}} \ket{ii} \in \mathcal{H}_\mathcal{S}\otimes \mathcal{H}_\mathcal{S}$:
\begin{gather}
\label{eqn::Choi}
    \Lambda = (\hat \Lambda \otimes \mathcal{I})\left[\ketbra{\Phi^+}{\Phi^+}\right] = \sum_{i,j=1}^{d_\mathcal{S}} \hat \Lambda[\ketbra{i}{j}] \otimes \ketbra{i}{j}\, ,
\end{gather}
where $\mathcal{I}$ is an identity map on $\mathcal{B}(\mathcal{H}_\mathcal{S})$. We denote maps with a caret, and their respective Choi matrices by the same letter without a caret. Whenever there is no risk of confusion, we will drop the explicit distinction between a map and its Choi state. For ease of notation, here, we explicitly distinguish between the input and output space of $\hat \Lambda$, \textit{i.e.}, $\hat \Lambda: \mathcal{B}(\mathcal{H}^{\mathrm{in}}_\mathcal{S}) \rightarrow \mathcal{B}(\mathcal{H}^{\mathrm{out}}_\mathcal{S})$ and $\Lambda \in \mathcal{B}(\mathcal{H}^{\mathrm{out}}_\mathcal{S}) \otimes \mathcal{B}(\mathcal{H}^{\mathrm{in}}_\mathcal{S})$~\footnote{In general, the map $\hat \Lambda$ could be a mapping between spaces of different size~\cite{chiribella_transforming_2008,milz_introduction_2017}. Here, we merely distinguish between the spaces to clarify our notation.}.  With this, the action of $\hat \Lambda$ on a state $\rho\in \mathcal{B}(\mathcal{H}^{\mathrm{in}}_\mathcal{S})$ can be written as
\begin{gather}
\label{eqn::ChoiAction}
    \hat \Lambda[\rho] = \tr_\mathrm{in}\left[(\mathbbm{1}_\mathrm{out}\otimes \rho^{\mathrm{T}})\Lambda\right]\, ,
\end{gather}
where $\tr_\mathrm{in}$ signifies a trace over $\mathcal{H}_{\mathcal{S}}^{\mathrm{in}}$, $\cdot^{\mathrm{T}}$ denotes the transpose in a fixed basis, and $\mathbbm{1}_\mathrm{out}$ is an identity matrix on $\mathcal{H}_\mathcal{S}^{\mathrm{out}}$~\cite{chiribella_transforming_2008,milz_introduction_2017}.

Using the Choi representation~\eqref{eqn::Choi}, the matrix $\Lambda$ can be experimentally reconstructed by measuring the output states $\left\{\rho_\mu'\right\}_{\mu=1}^{d_\mathcal{S}^2}$ for a basis of input states. It is given by
\begin{gather}
 \label{eqn::Channel}
 \Lambda = \sum_{\alpha=1}^{d_\mathcal{S}^2} \rho'_\mu \otimes \omega_\mu^{\text{T}},
\end{gather}
where $\tr\left[\rho_\mu\omega_\nu\right] = \delta_{\mu\nu}$. The set $\left\{\omega_\nu\right\}$ is called the dual set to the basis $\left\{\rho_\mu\right\}$. It can be constructed for any set of linearly independent matrices. By insertion into Eq.~\eqref{eqn::ChoiAction} it can be seen that the channel $\Lambda$ reconstructed in this way yields the correct output state $\rho_\mu'$ for each input $\rho_\mu$, and hence, by linearity, for any state $\rho \in \mathcal{B}\left(\mathcal{H}_\mathcal{S}\right)$; it is a valid description of a (single step) process whenever the preparation of system states is independent of the environment. The fact that measuring the outputs for a basis of inputs yields a correct dynamical description directly generalizes to more complex experimental scenarios.

The operational generalization of $\hat{\Lambda}$ to multiple times leads to the process tensor formalism~\cite{pollock_complete_pra,pollock_operational_prl}. Consider an experimenter performing control operations on the system at various times. These control operations could, \textit{e.g.}, be unitary operations or sequential measurements. In the most general setting, the experimenter could, at times $t_0,\dots,t_{N-1}$, act with CP maps $\hat{\mathcal{A}}_{\mu_0}, \dots, \hat{\mathcal{A}}_{\mu_{N-1}}$ (which we will assume, henceforth, can be implemented on a time scale much smaller than typical time scales of the system dynamics) on the system and measure the resulting final state at $t_N$. The process tensor $\hat{\mathcal{T}}^{N:0}$ is defined as the linear mapping from the control operations made by the experimenter to the final state: $\hat{\mathcal{T}}^{N:0}[\hat{\mathcal{A}}_{\mu_0}, \dots, \hat{\mathcal{A}}_{\mu_{N-1}}] = p(\hat{\mathcal{A}}_{\mu_0}, \dots, \hat{\mathcal{A}}_{\mu_{N-1}}) \rho(\hat{\mathcal{A}}_{\mu_0}, \dots, \hat{\mathcal{A}}_{\mu_{N-1}})$. The sequence of control operations might not be applied deterministically and therefore need not preserve trace. Consequently the final output $\rho$ is observed with probability $p$, and both are functions of the control operations.

As the operations of this sequence are chosen independently, the corresponding Choi matrix is given by $\mathcal{A}_{\mu_0} \otimes \cdots \otimes \mathcal{A}_{\mu_{N-1}}$. Measuring the output states for a full basis of sequences of CP operations then allows one -- just like in the case of channels $\hat\Lambda$ -- to reconstruct a matrix $\mathcal{T}^{N:0}$ that fully describes the underlying process. In detail, we have  
\begin{gather}
 \label{eqn::ProcTensRec}
 \mathcal{T}^{N:0} = \sum_{\vec \mu} \rho'_{\vec \mu} \otimes \Delta_{\vec \mu}^\text{T},
\end{gather}
where  $\rho'_{\vec \mu}$ is the (non-unit trace) output state for the choice of CP operations $\mathcal{A}_{\vec \mu} = \bigotimes_{k=0}^{N-1} \mathcal{A}_{\mu_k}$ at times $t_0,\dots t_{N-1}$, respectively. $\left\{\mathcal{A}_{\vec \mu}\right\}$ forms a basis of the space of $N$-sequences of CP operations and $\left\{\Delta_{\vec \mu}\right\}$ is its dual set, \textit{i.e.}, $\tr\left[\mathcal{A}_{\vec \mu} \Delta_{\vec \nu} \right] = \delta_{\vec \mu \, \vec \nu}$. The action of the resulting \textit{process tensor} $\mathcal{T}^{N:0}$ on a general sequence $\mathbf{A}_{N-1:0} = \sum_{\vec \mu} s_{\vec \mu} \mathcal{A}_{\vec \mu}$ of $N$ (possibly temporally correlated ~\cite{pollock_complete_pra}) CP operations is given by
\begin{align}
\notag
    \hat{\mathcal{T}}^{N:0}[\hat{\mathbf{A}}_{N-1:0}] &= \tr_{\mathrm{in}}\left[(\mathbbm{1}_\mathrm{out} \otimes \mathbf{A}^{\mathrm{T}}_{N-1:0})\mathcal{T}^{N:0} \right] \\ &:= \rho'(\mathbf{A}_{N-1:0})\, ,
    \label{eqn::ChoiActionProc}
\end{align}
where $\tr_\mathrm{in}$ is a trace over the Hilbert spaces that $\mathbf{A}_{N-1:0}$ acts on, while $\mathbbm{1}_\mathrm{out}$ is the identity matrix on the Hilbert space of the resulting final state~\cite{milz_introduction_2017}. As for the case of channels, it can be shown by insertion that Eq.~\eqref{eqn::ChoiActionProc} yields the correct output state for every basis sequence $\mathcal{A}_{\vec \mu}$, and consequently, by linearity, it yields the correct output state $\rho'(\mathbf{A}_{N-1:0})$ for \textit{any} sequence $\mathbf{A}_{N-1:0}$ of operations that are implemented on the system at times $t_0,\dots, t_{N-1}$. In this sense, each of the basis sequences can be considered as a particular \textit{trajectory} of the underlying process, and the set of basis trajectories allows for the construction of all other trajectories~\cite{sakuldee_non-markovian_2018}. For example, if the experimenter steered the system of interest to a desired final state $\rho'$ by acting on it with unitary operations $\hat{\mathcal{V}}_0,\dots,\hat{\mathcal{V}}_{N-1}$ at times $t_0,\dots,t_{N-1}$, the resulting final state would be given by $\rho' = \tr_{\mathrm{in}}[(\mathbbm{1}_{\mathrm{out}}\otimes \mathcal{V}^{\mathrm{T}}_0 \otimes \cdots \mathcal{V}^\mathrm{T}_{N-1})\mathcal{T}^{N:0}]$. 

\begin{figure}
\centering
\includegraphics[width=0.9\linewidth]{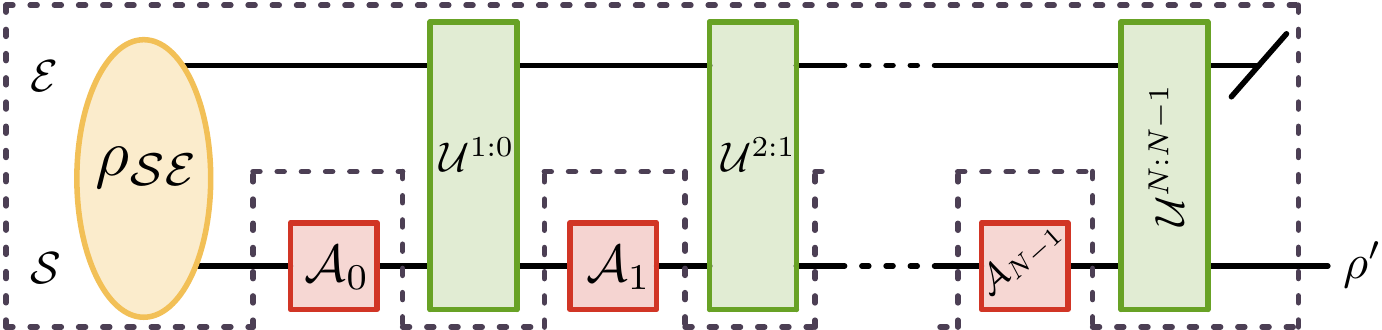}
\caption{\textit{Depiction of a multi-step process.} The local operations $\mathcal{A}_k$ act at times $t_k$. In between, the dynamics is governed by the system-environment unitaries $\mathcal{U}_{k+1:k}$. The final state $\rho' = \rho'(\mathbf{A}_{N-1:0})$ is determined by QST at the time $t_N$. The corresponding process tensor $\mathcal{T}^{N:0}$ is depicted by the dotted line.}
\label{fig::ProcTens}
\end{figure}

The process tensor is a completely positive map that satisfies trace conditions that ensure its causal ordering~\cite{chiribella_transforming_2008, chiribella_theoretical_2009, pollock_complete_pra}, and it can be shown that every process tensor has a circuit representation that consists of an initial (possibly correlated) system environment state and unitary system-environment dynamics between the local operations performed by the experimenter (see Fig.~\ref{fig::ProcTens}). It can be reconstructed by probing the system with complete set of operations at each time step. Consequently, the process tensor is the most general description of (non-Markovian) open quantum system dynamics and the traditional description of open system dynamics in terms of channels can be recovered from one-step process tensors -- also called \textit{superchannels} -- under the assumption of an initial system-environment state of product form~\cite{modi_operational_2012}.

\section{Restricted process Tensors}
\label{sec::Restr}
As mentioned above, the full reconstruction of quantum dynamics requires performing a basis set of CP operations at each time $t_0,\dots,t_{N-1}$. This implies an exponentially large number of necessary experiments. Moreover, the  vast majority of real-world experiments are necessarily, or by design, constrained in the amount of control that the experimenter can exert on the system of interest. For example, as already mentioned, both optical experiments and those concerned with quantum control, as well as IBM's publicly available quantum computer only admit for intermediate unitary operations. Given that unitary operations -- and generally any limited set of operations -- do not form a linear vector space, it is \textit{a priori} unclear how a meaningful description of an experiment can be reconstructed in a finite number of experiments. The same holds true for the analysis of initial correlations and memory effects (see Secs.~\ref{subsec::InitCorrWit} and~\ref{subsec::NonMark}). In this section, we show how to reconstruct operationally meaningful \textit{restricted} process tensors for any kind of experimental control and work out this reconstruction for the cases where only unitary operations and projective measurements can be performed, respectively.

\subsection{Restricted tomographic reconstruction}
\label{subsec::Restr}
We denote by $\mathcal{F}$ the set of available operations that can be performed by the experimenter.  For example, $\mathcal{F}$ could be the set of unitary operations acting on the system, or a set of operations generated by a finite control algebra~\cite{viola_dynamical_1999}. Let $W = \text{Span}\left(\mathcal{F}\right)$ be the vector space spanned by this set, \textit{i.e.}, the vector space that consists of linear combinations of elements of $\mathcal{F}$. While a restricted set $\mathcal{F}$ is in general not sufficient to reconstruct the full process tensor (\textit{i.e.}, the available operations are in general not sufficient to obtain all possible multi-step correlation functions), the experimental reconstruction of a process tensor based on the set $\mathcal{F}$ follows directly from Eq.~\eqref{eqn::ProcTensRec}: It can be reconstructed by measuring the output states corresponding to a \textit{basis} of operations, which, in this case, is the basis of the space $W$ that $\mathcal{F}$ spans. 

We begin by noting that there exists a set  $\left\{f_\alpha\right\}^{d_w-1}_{\alpha=0}\subset \mathcal{F}$ that constitutes a basis of $W$, where $d_w$ is the dimension of $W$. Following Eq.~\eqref{eqn::ProcTensRec}, the restricted process tensor $\mathcal{T}_\mathcal{F}^{N:0}$ is then given by 
\begin{gather}
 \label{eqn::partProc}
 \mathcal{T}^{N:0}_{\mathcal{F}} = \sum_{\vec \alpha} \eta'_{\vec \alpha} \otimes \Theta_{\vec \alpha}^{\text{T}}, 
\end{gather}
where $\eta'_{\vec \alpha}$ is the output state corresponding to the sequence of operations $f_{\vec \alpha} = \bigotimes_{\vec \alpha} f_{\alpha_k}$ and  $\tr\left(\Theta_{\vec \gamma} f_{\vec \alpha} \right) = \delta_{\vec \alpha \, \vec \gamma}$. Due to linearity, the resulting restricted process tensor yields the correct output state for any (possibly temporally correlated~\cite{pollock_complete_pra}) admissible sequence $\mathbf{A}_{N-1:0}$ that lies in $W^{\otimes N}$. The sequence $\mathbf{A}_{N-1:0}=\sum_{\vec \alpha} b_{\vec \alpha} f_{\vec \alpha}$, with $b_{\vec \alpha}\in \mathbb{R}$ can be written as a multilinear combination of basis elements $f_\alpha$. Hence, we have
\begin{align}
\notag
\hat{\mathcal{T}}^{N:0}_{\mathcal{F}}\left[\hat{\mathbf{A}}_{N-1:0}\right] &= \sum_{\vec \alpha} b_{\vec \alpha} \tr_{\mathrm{in}}[ (\mathbbm{1}_\mathrm{out} \otimes f_{\vec \alpha}^{\mathrm{T}})\mathcal{T}^{N:0}_{\mathcal{F}}] \\ &=  \sum_{\vec \alpha} b_{\vec \alpha} \, \eta'_{\vec \alpha}\,, 
\end{align}
where we have used that $\mathcal{T}_{\mathcal{T}}^{N:0}$ maps every basis element $f_{\vec\alpha}$ to its correct output state. We emphasize that, in general, $W$ contains a larger set of CP maps than $\mathcal{F}$. For example, the space spanned by the set of unitary maps contains all unital maps (see Sec.~\ref{sec:unitaries} below), and a restricted process tensor reconstructed from a set of unitary operations can predict the output state for any sequence of unital maps.

Eq.~\eqref{eqn::partProc} establishes how a process tensor for an $N$-step process can be reconstructed based on a restricted set $\mathcal{F}$ of local operations; given $\mathcal{F}$, one derives the dimension $d_w$ of the vector space that $\mathcal{F}$ spans, determines a set $\left\{f_{\alpha}\right\}_{\alpha=0}^{d_w-1}\subset \mathcal{F}$ of $d_w$ linearly independent operations, and measures the output state for each of the $d_w^N$ possible $f_{\vec \alpha}$. The reconstructed process tensor yields the correct output state for any admissible sequence of operations $\mathbf{A}_{N-1:0} \in W^{\otimes N}$. 

Unsurprisingly, outside the space $W$, the restricted process tensor does not yield meaningful results; the vector space of all possible sequences of local operations at $N$ time steps can be decomposed as $\left[W \oplus W^{\perp}\right]^{\otimes N}$, where $W^{\perp}$ is the orthogonal complement of $W$. Denoting the basis of $\left\{W^{\perp}\right\}$ by $\{g_\beta\}_{\beta = 0}^{d^4_\mathcal{S}-d_w-1}$, the set of all possible $N$-fold tensor products of elements of $\left\{f_\alpha\right\}$ and $\{g_\beta\}$ forms a basis of $\left[W \oplus W^{\perp}\right]^{\otimes N}$ and it is easy to see that every basis element that does not exclusively contain elements of $\left\{f_\alpha\right\}$ gets mapped to zero by $\mathcal{T}_\mathcal{F}^{N:0}$. Consequently, the action of $\mathcal{T}_\mathcal{F}^{N:0}$ and $\mathcal{T}^{N:0}$ coincide on $W^{\otimes N}$, but the restricted process tensor does not allow for meaningful predictions of the output state for any sequence of preparations $\mathbf{A}_{N-1:0} \notin W^{\otimes N}$. 

By how much the full and the restricted process tensor differ depends both on the number of basis elements that get mapped onto zero by $\mathcal{T}_{\mathcal{F}}^{N:0}$, as well as the action of the full process tensor on said basis elements. The number is given by $d_{\mathcal{S}}^{4N}-d_w^N$, while the action of $\mathcal{T}^{N:0}$ outside of $W^{\otimes N}$ depends on the system-environment unitary maps that govern the evolution in between the time steps $t_k$. 

We emphasize that the restricted process tensor framework is explicitly tailored to enable the description of non-Markovian processes; for Markovian processes, the dynamics between any two time steps $t_k$ and $t_{k+1}$ is given by a CPTP map $\Lambda_{k+1:k}$~\cite{pollock_complete_pra,pollock_operational_prl} that could be reconstructed with limited control (for example, if the available operations allow for state tomography at each time step). However, as soon as memory effects play a non-negligible role, the only meaningful description of the underlying process is given by the restricted process tensor.

\subsection{Projective measurements and unitary control}
\label{subsec::UnitProj}

Here, we examine two extremal cases of experimental control -- the case where only unitary operations are available, and the case where only projective measurements can be performed -- and illustrate the reconstruction and domain of applicability of restricted process tensors. The two considered cases are extremal in the sense that unitary operations do not allow one to infer any information about the state of the system, while projective measurements provide direct access to information, but lead to collapse into a definite (pure) state, and destroy system-environment correlations. As they decouple the system from its environment, projective measurements enable the construction of direct witnesses of correlations (see Sec.~\ref{subsec::InitCorrWit}). These two sets of operations are mathematically tractable enough to derive the dimensions of their respective spans. We emphasize that being able to perform both unitary operations as well as measurements is tantamount to spanning the space of all control operations.

\subsubsection{Unitary operations}
\label{sec:unitaries}

A unitary map $\hat{\mathcal{V}}$ acting on a quantum state $\rho \in \mathcal{B}\left(\mathcal{H}_{\mathcal{S}}\right)$ is of the form $\hat{\mathcal{V}}\left[\rho\right] = V\rho \, V^{\dagger}$, where $V \in \text{SU}\left(d_\mathcal{S}\right)$ is a unitary matrix. From Eq.~\eqref{eqn::Choi}, it is straightforward to see that, up to normalization, a unitary map $\hat{ \mathcal{V}}$ corresponds to a pure, maximally entangled state $\mathcal{V} \in \mathcal{B}\left(\mathcal{H}_{\mathcal{S}}\right) \otimes \mathcal{B}\left(\mathcal{H}_{\mathcal{S}}\right)$. Denoting the span of the set of unitary maps by $W_{\mathcal{U}}$, we see that any element $\mathcal{V} \in W_{\mathcal{U}}$ can be written as
\begin{gather}
\label{eqn::SpanUnit}
\mathcal{V}  = \sum_{\mu} b_\mu\left(\mathbbm{1} \otimes \mathbbm{1} + \sum_{k,l=1}^{d_{\mathcal{S}}^2-1}c^{(\mu)}_{kl} \sigma_k \otimes \sigma_l\right), 
\end{gather}
where $\mathbbm{1}$ is the identity matrix on $\mathcal{H}_\mathcal{S}$ and $\left\{\sigma_k\right\}$ are the traceless generators of $\text{SU}\left(d_\mathcal{S}\right)$. The converse is also true; any map that can be expressed in the form of Eq.~\eqref{eqn::SpanUnit} lies in $W_\mathcal{U}$~\cite{poon_preservers_2015}. The set of completely positive maps contained in $W$ coincides with the set of unital maps (those that leave the completely mixed state invariant)~\cite{mendl_unital_2009}. 

From the fact that the operators $\left\{\mathbbm{1} \otimes \mathbbm{1}, \sigma_k \otimes \sigma_l\right\}_{k,l=1}^{d_{\mathcal{S}}^2 - 1}$ are linearly independent, we deduce that $W_\mathcal{U}$ is $d_u = \left(d_{\mathcal{S}}^2-1\right)^2+1$ dimensional (whereas the space of all possible local operations is $d_{\mathcal{S}}^4$-dimensional). In the qubit case, for example, we have $d_u=10$ and a basis of $W_\mathcal{U}$ that consists of unitary maps can be readily constructed (see App.~\ref{app::BasisUnitQubit}; note that the maps $\left\{\mathbbm{1} \otimes \mathbbm{1}, \sigma_k \otimes \sigma_l\right\}$ also form a basis of $W$, but do not all correspond to unitary maps). In the general, higher dimensional case, it is sufficient to randomly choose a set of $d_u$ linearly independent unitary maps for the construction of the restricted process tensor. 

A process tensor $\mathcal{T}^{N:0}_\mathcal{U}$ constructed based on the set of unitary local operations can be meaningfully applied to any sequence $\mathbf{D}_{N-1:0} \in W_\mathcal{U}^{\otimes N} $of (possibly temporarily correlated) unital maps. This means that, by measuring the output states for $[\left(d_{\mathcal{S}}^2-1\right)^2+1]^N$ sequences of independent unitary operations, the output state for any sequence of unital maps can be predicted.

\subsubsection{Projective operations}
\label{subsec::Proj}

If the experimental setup only allows for projective measurements of the system of interest, the set  $\mathcal{F}_\mathcal{P}$ of available operations coincides with rank-$1$ projections. We denote the span of $\mathcal{F}_\mathcal{P}$ by $W_{\mathcal{P}}$.  The action of a map $\hat{\mathcal{Q}} \in \mathcal{F}_\mathcal{P}$ on a quantum state $\rho \in \mathcal{B}(\mathcal{H}_\mathcal{S})$ is given by 
\begin{gather}
 \label{eqn::projMeas}
 \rho' = \hat{\mathcal{Q}}[\rho] = Q \rho\, Q = Q\tr \left(Q \rho\right) ,
\end{gather}
where $Q \in \mathcal{B}\left(\mathcal{H}_\mathcal{S}\right)$ is a pure state, \textit{i.e.}, $Q = \ketbra{q}{q}$. The state $\rho'$ after the action of $\hat{\mathcal{Q}}$ is given by $Q$, and $\tr(\rho')$ yields the probability to measure the outcome corresponding to $Q$. We emphasize that projective measurements destroy any correlations between the system and its environment, as $\hat{\mathcal{Q}}\otimes \mathcal{I}_\mathcal{E}[\rho_{\mathcal{SE}}] = Q\otimes \tr_\mathcal{S}[(Q\otimes \mathbbm{1}_\mathcal{E})\rho_\mathcal{SE}]$, where $\rho_{\mathcal{SE}} \in \mathcal{B}(\mathcal{H}_\mathcal{S}) \otimes \mathcal{B}(\mathcal{H}_\mathcal{E})$ is a system-environment state and $\mathcal{I}_\mathcal{E}$ is the identity map on $\mathcal{B}(\mathcal{H}_\mathcal{E})$. Consequently, they can be used to construct witnesses for system-environment correlations (see Sec.~\ref{subsec::InitCorrWit} and Ref.~\cite{kuah_how_2007}).

The Choi state of a map $\hat{\mathcal{Q}} \in \mathcal{F}_{\mathcal{P}}$ is of the form $Q \otimes Q^{\text{T}}$; therefore, the Choi state of any map $\hat{\mathcal{N}} \in W_{\mathcal{P}}$ must be of the form 
\begin{gather}
 \label{eqn::ProjSpan}
 \mathcal{N} = \sum_\nu b_\nu \, Q_\nu \otimes Q_\nu^{\text{T}}\, ,
\end{gather}
where $Q_\nu\in \mathcal{B}(\mathcal{H}_\mathcal{S})$ are pure states and $b_\nu \in \mathbb{R}$. $W_{\mathcal{P}}$ is at most $\frac{1}{4}d_\mathcal{S}^2(d_\mathcal{S}+1)^2$ dimensional (see App.~\ref{app::Proj}). For the qubit case, a set $\left\{Q_\nu\right\}$ of $\frac{1}{4}d_\mathcal{S}^2(d_\mathcal{S}+1)^2 = 9$ pure states that correspond to linearly independent maps $\{\mathcal{Q}_\nu\}$ has been constructed in Ref.~\cite{kuah_how_2007} (and is reproduced here in App.~\ref{app::Proj}).

A one-step process tensor $\mathcal{T}^{1:0}_\mathcal{P}$ constructed based on projections alone can be reconstructed by measuring the output states for $\frac{1}{4}d_\mathcal{S}^2(d_\mathcal{S}+1)^2$ linearly independent projections $\mathcal{Q} \in \mathcal{F}_{\mathcal{P}}$; it can meaningfully be applied to any CP map $\mathcal{N}$ that satisfies $\tr_{\mathrm{out}}(\mathcal{N}) = 
(\tr_{\mathrm{in}}(\mathcal{N}))^{\text{T}}$, where $\tr_{\mathrm{in}}$ and $\tr_{\mathrm{out}}$ denote the trace with respect to the first and second subsystem in Eq.~\eqref{eqn::ProjSpan}, respectively. Analogously, an $N$-step restricted process tensor $\mathcal{T}_\mathcal{P}^{N:0}$ can be reconstructed by measuriung the output states for $\frac{1}{4}[d_\mathcal{S}^2(d_\mathcal{S}+1)^2]^N$ linearly independent sequences of projections; it can be applied to any physically admissible sequence of operations in $W_{\mathcal{P}}^{\otimes N}$.

\subsection{Properties of restricted process tensors}
\label{subsec::RestrProp}
The full description of an $N$-step process, \textit{i.e.}, $\mathcal{T}^{N:0}$, is a linear CP map. It also displays a `containment property'~\cite{pollock_complete_pra}; given this full description, the dynamics of the process for a smaller subset of time steps can be obtained. For example, given a process tensor $\mathcal{T}^{4:0}$, that describes a process where the system can be manipulated at time-steps $t_0,t_1,t_2$ and $t_3$, it is possible to predict the outcome state at $t_3$ if the system was only manipulated at times $t_1$ and $t_2$ (but not at $t_0$), \textit{i.e.}, $\mathcal{T}^{4:0}$ \textit{contains} the correct `smaller' process tensor $\mathcal{T}^{2:1}$. Concretely, this statement implies that the information about $N$-step correlation functions contains all the information about correlation functions for fewer time steps. It turns out, that restricted process tensors are in general not CP, and while they display a \textit{partial} containment property, this containment property is in general qualitatively different from the one displayed by full process tensors. 

\subsubsection{Complete positivity}
By construction, the restricted process tensor $\mathcal{T}^{N:0}_{\mathcal{F}}$ is a linear map. However, it is \textit{a priori} unclear whether the remaining properties of the full process tensor also hold in the restricted case. 

To see more clearly what complete positivity means for process tensors, consider a set of control operations acting on both the system and an ancilla $\mathcal{X}$. The system is undergoing a quantum process while the ancilla is simply sitting on the side. Mathematically this process can be written as
\begin{gather}
\label{eq:restrictedcp}
\hat{\mathcal{T}}^{N:0} \otimes \mathcal{I}_{\mathcal{X}} [\hat{\mathbf{A}}_{N-1:0}^{\mathcal{SX}}] = \hat{\Gamma}_{\mathcal{X} \rightarrow \mathcal{SX}}, 
\end{gather}
where $\hat{\Gamma}_{\mathcal{X} \rightarrow \mathcal{SX}}$ is a mapping from $\mathcal{B}(\mathcal{H}_\mathcal{X})$ to $\mathcal{B}(\mathcal{H}_\mathcal{S})\otimes \mathcal{B}(\mathcal{H}_\mathcal{X})$.  A process tensor is completely positive, iff $\Gamma_{\mathcal{X} \rightarrow \mathcal{SX}}$ is positive (\textit{i.e.}, $\hat{\Gamma}_{\mathcal{X} \rightarrow \mathcal{SX}}$ is completely positive) for any CP map $\hat{\mathbf{A}}_{N-1:0}^{\mathcal{SX}}$ and any ancilla $\mathcal{X}$. Conversely, if there exists a physical $\mathcal{SX}$ control operation that leads to a map in Eq.~\eqref{eq:restrictedcp} that is not CP, then the process tensor is not completely positive. For restricted process tensors $\mathcal{T}_\mathcal{F}^{N:0}$, complete positivity is not a well-defined property. By construction, they can only be meaningfully applied to a subset of control operations, while they yield physically non-sensical results for operations outside this subset. Put differently, there are control operations $\hat{\mathbf{A}}_{N-1:0}^{\mathcal{SX}}$ in Eq.~\eqref{eq:restrictedcp} whose restriction on $\mathcal{S}$ lies outside the set of operations that $\mathcal{T}_\mathcal{F}^{N:0}$ is defined on. We emphasize that this is also true if the Choi state $\mathcal{T}^{N:0}_\mathcal{F}$ of the restricted process tensor was coincidentally positive. This can, \textit{e.g.}, happen when only projective measurements in a fixed basis can be performed which leads to positive duals in Eq.~\eqref{eqn::partProc}. 

In general, restricted process tensors $\mathcal{T}_\mathcal{F}^{N:0}$ reconstructed according to Eq.~\eqref{eqn::partProc} are not positive matrices, but can always be extended to a positive matrix (that is, $\mathcal{T}^{N:0}$). Nonetheless, they are -- unlike non-CP channels~\cite{shaji_whos_2005, milz_introduction_2017} -- operationally well-defined and the `break-down' of complete positivity is not fundamental but merely due to the description of the process we chose. This situation is similar to incomplete tomography of a quantum state; the probabilities obtained from probing a state with a POVM that is not informationally complete might be faithfully reproduced by a non-positive `density matrix'. This matrix would yield correct probabilities for each of the POVM elements used to probe the state, but would not contain any information about probabilities for other POVM elements. Quite obviously, the non-positivity of a density matrix reconstructed in this way is neither fundamental, nor does it imply negative probabilities, but is simply reminiscent of the representation we chose.

\subsubsection{Containment}

The containment property for the process tensor turns out to be fundamentally important. As mentioned above, formally, it states that for any $0\leq j < k \leq N$, the correct process tensor $\mathcal{T}^{k:j}$ can be obtained from $\mathcal{T}^{N:0}$. The containment property can be thought of as a causality conditions for the process and it has recently been employed to generalize the Kolmogorov extension theorem to the quantum case~\cite{milz_kolmogorov_2017}.

In general, a restricted set of operations does not allow one to deduce the state of the system at intermediate times (see below), and as such -- except for special cases (see App.~\ref{app::ICPOVM}) -- $\Tcal^{N:0}_\mathcal{F}$ does not enable the derivation of intermediate restricted process tensors $\Tcal^{k:j}_\mathcal{F}$. However, it enables the calculation of \textit{probabilities} for subsets of time steps, and thus, partial containment property for the restricted process tensor exists. In detail, this means that we can obtain any sub-process $\mathcal{M}^{k:j}_{\mathcal{F}}$ from a given restricted process tensor $\mathcal{T}^{N:0}_{\mathcal{F}}$ that yields the correct probabilities for the occurence of CP maps $\mathcal{A}_j,\dots,\mathcal{A}_{k-1}$ at times $t_j,\dots,t_{k-1}$, but does not yield an output state.

We begin the construction of $\mathcal{M}^{k:j}_{\mathcal{F}}$ by inserting deterministic (CPTP) operations $\mathbf{A}_{N-1:k}$ for all time steps greater than $k-1$ and tracing over the final state at $t_N$ in Eq.~\eqref{eqn::partProc} to get
\begin{gather}
\hat{\mathcal{M}}^{k:0}_\mathcal{F}[\cdot] :=
\tr\left(\hat{\mathcal{T}}^{N:0}_\mathcal{F}[\hat{\mathbf{A}}_{N-1:k}, \cdot]\right)\, ,
\end{gather}
where the Choi matrix of $\hat{\mathcal{M}}^{k:0}_\mathcal{F}$ is given by $\mathcal{M}^{k:0}_\mathcal{F} = \tr_{N:k}\left[(\openone_N \otimes \mathbf{A}_{N:k}^\mathrm{T} \otimes \openone_{k-1:0}) \mathcal{T}^{N:0}_\mathcal{F}  \right]$. The CPTP operations $\mathbf{A}_{N-1:k}$ can be chosen arbitrarily, as long as the action of $\mathcal{T}_\mathcal{F}^{N:0}$ is well-defined on them. It is important to note that, unlike $\mathcal{T}_\mathcal{F}^{k:0}$, the map ${\mathcal{M}}^{k:0}_\mathcal{F}$ yields the correct probability of occurrence when applied to a sequence of CP maps, instead of the correct output state at $t_k$.

Next, for the time steps $0$ to $j-1$, we apply the `do nothing operation'. Formally, this means to `act' on the system with the identity operators $\mathcal{I}^{i}_\mathcal{S}$ at the time steps $t_{i}$ ($i<j$):
\begin{gather}
\label{eqn::ContRest}
 \hat{\mathcal{M}}_\mathcal{F}^{k:j}[\cdot] = \hat{\mathcal{M}}_\mathcal{F}^{k:0}[\, \cdot \, , \mathcal{I}^\mathcal{S}_{j-1}, \dots, \mathcal{I}^\mathcal{S}_0]\,.
\end{gather}
Here we assumed that the experimental setup allows for the `do nothing' map. If this is not the case, we can replace the identity map with the default map for the process and obtain the corresponding sub-process. Again, it is important to clearly delineate between the object $\mathcal{M}_\mathcal{F}^{k:j}$ that we constructed, and the `actual' restricted process tensor $\mathcal{T}_\mathcal{F}^{k:j}$ for the respective time steps. $\mathcal{T}_\mathcal{F}^{k:j}$ yields the correct output \textit{state}, while $\mathcal{M}_\mathcal{F}^{k:j}$ yields the correct \textit{probability} for the occurence of any sequence of CP maps that can be implemented at the respective time steps. The two maps are directly related, though, via $\hat{\mathcal{M}}_\mathcal{F}^{k:j} = \tr \circ \hat{\mathcal{T}}_\mathcal{F}^{k:j}$, where $\circ$ denotes the sequential composition.  While $\mathcal{M}_\mathcal{F}^{k:j}$ can \textit{always} be obtained from $\mathcal{T}_\mathcal{F}^{N:0}$, it is generally not possible to calculate $\mathcal{T}_\mathcal{F}^{k:j}$ given only $\mathcal{T}_\mathcal{F}^{N:0}$.

Generally, the restricted process tensor does not allow one to determine the state of the system at any intermediary time. For example, if the operations that the experimenter can implement are unitary, no information about the system is obtained when the operations $\mathcal{V}_0, \dots, \mathcal{V}_{N-1}$ are performed at the times $t_0,\dots,t_{N-1}$, and a restricted process tensor that was reconstructed for sequences of unitary operations would not allow one to infer the state of the system at any time $t_k \neq t_{N}$. Nonetheless, $\mathcal{M}_\mathcal{F}^{k:j}$ constructed according to Eq.~\eqref{eqn::ContRest} yields the correct probability to obtain any  non-deterministic sequence of CP operations $\mathbf{A}_{k-1:j}$ that the restricted process tensor can meaningfully act on, and is thus a meaningful descriptor for this subset of times. In this well-defined sense, restricted process tensors possess a partial containment property. 

On the other hand, if the set of operations $\mathcal{F}$ that is available to the experimenter allows them to perform an informationally complete measurement (see, \textit{e.g.},~\cite{bengtsson_geometry_2007}, chapter 10) at each time, it is possible to derive `actual' intermediate restricted process tensors $\hat{\mathcal{T}}_{\mathcal{F}}^{k:j}$ from $\hat{\mathcal{T}}_{\mathcal{F}}^{N:0}$, that allow one to not only obtain the correct outcome probabilities, but also the final state at $t_{k}$; availability of an informationally complete measurement implies that the state of the system at each time can be inferred by simple post-processing of the data contained in $\hat{\mathcal{T}}_{\mathcal{F}}^{N:0}$ and consequently all intermediate restricted process tensors can be constructed (see App.~\ref{app::ICPOVM} for details). This is for example the case when the set of performable operations coincides with the set of projective measurements.

\section{Witnesses for memory effects}

The full process tensor contains all information that can be locally be inferred about the dynamics of a system. This means that it also contains all accessible information about correlations between the system and its environment, as well as memory effects in the process~\cite{modi_operational_2012, pollock_complete_pra}. In recent years, much effort has gone into developing a large number of \textit{witnesses}~\cite{rivas_entanglement_2010, rivas_quantum_2014, breuer_measure_2009} of system-environment correlations that have limited predictive power or assume a specific kind of experimental control~\cite{rodriguez-rosario_unification_2012, mazzola_dynamical_2012, laine_witness_2010}.

A word of caution is necessary before proceeding. The interpretation of results obtained from restricted control might lead to false positives for memory effects. To see this let us consider a simple example. Consider a single qubit subjected to a unitary process: two Hadamard gates. This is a perfectly Markovian process if we can apply arbitrary quantum operations to the qubit (at three times). However, suppose we are only allowed to measure the qubit in the computational ($z$) basis. In this case we conclude that the first channel due to a single Hadamard gate is maximally incoherent, while the overall process due to two Hadamard gates is the identity channel, and the process will be deemed non-Markovian. The apparent non-Markovianity is induced by the fact that we are not able to fully probe system. Therefore, detecting non-Markovianity and initial correlations with restricted control must be done with care. All of the witnesses we present are constructed in such a way that they cannot yield false positives for memory effects.

In what follows, we first show a way to construct operationally motivated correlation witnesses for any kind of experimental control, that make maximal use of the available operations. Subsequently, we extend these ideas to the construction of memory witnesses.  

\subsection{Witnesses for correlations from limited control}
\label{subsec::InitCorrWit}

A key feature of the process tensor is that it can describe the dynamics of a system that is initially correlated with its environment (see Fig.~\ref{fig::ProcTens}), whereas the conventional quantum channel formalism breaks down when the initial system-environment state is not of product form. Initial correlations are a generic feature of most experiments and represent a record of past system-environment interactions. Therefore, detecting initial correlations implies detecting non-Markovian dynamics (see Sec.~\ref{subsec::NonMark} and App.~\ref{App::ProofMark}). 

Consider an initial system-environment state (before preparation) of the form
\begin{gather}
 \rho_{\mathcal{SE}} = \rho_{\mathcal{S}} \otimes \rho_{\mathcal{E}} + \chi_{\mathcal{SE}},
\end{gather}
where $\rho_\mathcal{S} = \tr_\mathcal{E}\left[\rho_{\mathcal{SE}}\right]$, $\rho_\mathcal{E} = \tr_\mathcal{S}\left[\rho_{\mathcal{SE}}\right]$ and $\chi_{\mathcal{SE}}$ contains all initial correlations between the system of interest and its environment. In Ref.~\cite{modi_operational_2012} it is shown that from a reconstructed single step process the so-called correlation memory matrix $\mathcal{K}$ can be derived:
\begin{gather}
 \hat{\mathcal{K}}\left[\hat{\mathcal{A}}\right] = \tr_\mathcal{E}\left\{U \left[\left(\hat{\mathcal{A}} \otimes \mathcal{I}_\mathcal{E}\right) \left[\chi_{\mathcal{SE}}\right]\right]U^{\dagger} \right\}\, ,
\end{gather}
where $U$ is the system-environment unitary. This means that $\mathcal{K}$ encapsulates the time evolution of the initial correlations $\chi_{\mathcal{SE}}$. We emphasize that $\mathcal{K} \neq \mathbf{0}$ implies that the initial state $\rho_\mathcal{SE}$ was correlated, while $\mathcal{K}=\mathbf{0}$ merely implies that if initial correlations were present, the unitary $U$ does not allow for their detection via local operations. In this sense, $\mathcal{K}$ is the maximal information about correlations that can be inferred via local operations alone.

\begin{figure}
    \centering
    \includegraphics[width = 0.79\linewidth]{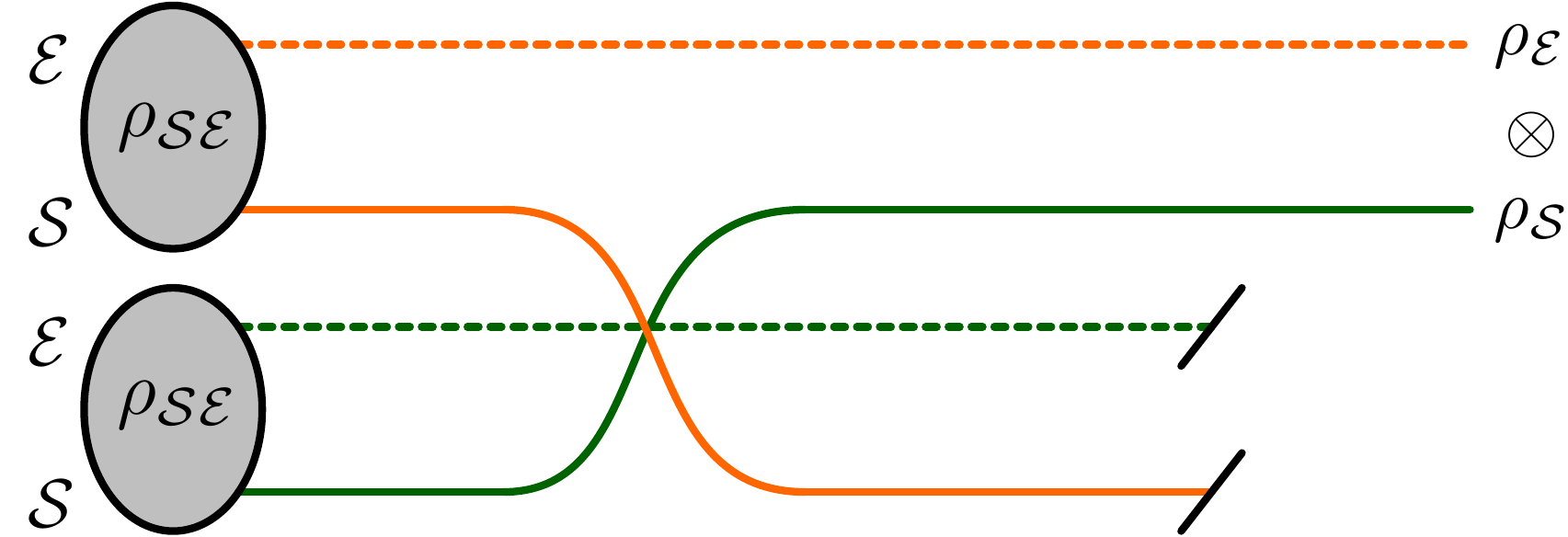}
    \caption{\textit{Preparation of a product state.} Two copies of the initial state $\rho_\mathcal{SE}$ are created and their system degrees of freedom are swapped by the operation $S^{(\mathcal{S})}_{12}$. Subsequently, the degrees of freedom of the second copy are disregarded.}
    \label{fig::SWAP}
\end{figure}

The derivation of the correlation memory matrix $\mathcal{K}$ from $\mathcal{T}^{1:0}$ relies on the fact that the experimenter has full control over the system of interest and its derivation is not directly applicable to  $\mathcal{T}^{1:0}_{\mathcal{F}}$. However, there are ways around this problem. Firstly, imagine an experimental situation that allows for the preparation of two copies of the initial state $\rho_{\mathcal{SE}}$, and admits a swap operation between the two states. In detail, let $S^{(\mathcal{S})}_{12}$ be a swap operation between the system Hilbert spaces of the two prepared initial states. With this, we can prepare an initial product state (see Fig.~\ref{fig::SWAP}):
\begin{gather}
 \label{eqn::SWAP}
 \tr_{\mathcal{S}_2\mathcal{E}_2}\left[S^{(\mathcal{S})}_{12}\left(\rho_{\mathcal{SE}} \otimes \rho_{\mathcal{SE}}\right)S^{(\mathcal{S})\dagger}_{12}\right] =  \rho_{\mathcal{S}} \otimes \rho_{\mathcal{E}}\, ,
\end{gather}
where $\tr_{\mathcal{S}_2\mathcal{E}_2}$ denotes a trace with respect to the second copy of $\rho_{\mathcal{SE}}$. For this initial product state, we can reconstruct the restricted one-step process tensor $\mathcal{L}^{1:0}_{\mathcal{F}}$, that describes the dynamics of the system for an initial product state. The corresponding correlation-memory matrix $\mathcal{K}_{\mathcal{F}} = \mathcal{T}^{1:0}_{\mathcal{F}} - \mathcal{L}^{1:0}_{\mathcal{F}}$ contains all the information about correlations that can be detected by local preparations $\mathcal{A}\in \text{Span}\left(\mathcal{F}\right)$. If $\mathcal{K}_{\mathcal{F}} \neq \mathbf{0}$, there exist initial system-environment correlations; if $\mathcal{K}_{\mathcal{F}} = \mathbf{0}$, no correlations that can be detected by local preparations in $\text{Span}\left(\mathcal{F}\right)$ are present. Depending on the dimension of $\text{Span}\left(\mathcal{F}\right)$ and the total dynamics $\mathcal{U}$, $\mathcal{K}_{\mathcal{F}}$ and $\mathcal{K}$ can contain the same information about the existence of initial correlations; in general, though, they will differ (see Sec.~\ref{sec:XTomo} for explicit examples).

In principle, the swap operation enables the construction of witnesses for correlations for any set $\mathcal{F}$ of available local operations, without requiring any knowledge of the initial system state $\rho_{\mathcal{S}}$ (unlike, \textit{e.g.}, the similar correlation witness proposed in~\cite{laine_witness_2010}). This construction of witnesses makes full use of the resources available to the experimenter, and $\mathcal{K}_{\mathcal{F}}$ is the maximum of information about correlations that can be inferred based on $\mathcal{F}$. However, its reconstruction requires both the preparation of two copies of the same initial state as well as the availability of a swap operation between the respective system degrees of freedom.

If the experimental setup does not allow for the direct reconstruction of $\mathcal{K}_{\mathcal{F}}$, witnesses for correlations can still be constructed depending on the set $\mathcal{F}$, namely if $\text{Span}(\mathcal{F})$ contains operations that decouple the system from its environment and map the system to a known state. For example, this is the case when the set of available operations contains projective measurements (see Sec.~\ref{subsec::Proj} and~\cite{kuah_how_2007}). 

Finally, if neither the swap $S^{(\mathcal{S})}_{12}$ nor local operations that decouple the system from its environment are available to the experimenter, then correlations and non-Markovianity may still be witnessed in a less direct sense: If the system is uncorrelated from its environment at some point, then its subsequent single-step dynamics is described by a CPTP channel, with a $d_\mathcal{S}^2$-dimensional input space. Therefore, if there are more than $d_\mathcal{S}^2$ linearly independent inputs (\textit{i.e.}, elements of $\text{Span}(\mathcal{F})$) to the restricted process tensor which are not consistent with a single CPTP channel, there must be correlations contributing to the dynamics.

\subsection{Witnesses for Non-Markovianity}
\label{subsec::NonMark}
If an experiment consists of more than one time-step, assertions about the existence of memory effects in the dynamics can be made. If the process displays memory effects, it is called non-Markovian, otherwise we call it Markovian. 

Classically, the dynamics of a system is Markovian if its time evolution at any time $t_{k-1}$ only depends on its current state, but not on its past. Recently, this concept has been generalized to the quantum case~\cite{pollock_complete_pra,pollock_operational_prl, li_concepts_2017}. Intuitively, the history dependence of a process at time $t_{k-1}$ can be probed by fixing its state at $t_{k-1}$ and analyzing its future evolution for different pasts. This intuition can be made manifest as follows: Let the experimenter perform any admissible sequence of operations $\mathbf{A}_{k-2:0}$ at times $t_0,\dots,t_{k-2}$. At time $t_{k-1}$ they measure the system and independently reprepare it in a fresh state. As this operation breaks the connection between the past and the future evolution on the level of the system, we call it a `causal break'. It can be expressed as $\mathcal{C}_{{k-1}}^{(ab)} = P^{(a)}_{k-1} \otimes \Pi_{k-1}^{(b)\mathrm{T}}$, where $\Pi_{k-1}^{(b)}$ is the POVM element corresponding to the measurement outcome, while $P^{(a)}_{k-1}$ is the freshly prepared state~\cite{pollock_complete_pra}. Just like in the classical case, a process has memory if the state of the system at the next time step $t_k$ does not only depend on the freshly prepared state $P^{(a)}_{k-1}$, but also on the measurement outcome at $t_{k-1}$ and/or the sequence of preparations $\mathbf{A}_{k-2:0}$ performed previously. Expressed in terms of the process tensor formalism, a process is Markovian iff
\begin{gather}
\label{eqn::Markov}
    \hat{\mathcal{T}}^{k:0}[\hat{\mathcal{C}}^{(ab)}_{k-1},\hat{\mathbf{A}}_{k-2:0}] \propto  \hat{\mathcal{T}}^{k:0}[\hat{\mathcal{C}}^{(ab')}_{k-1},\hat{\mathbf{A}}^\prime_{k-2:0}]\, ,
\end{gather}
$\forall a, b, b', \hat{\mathbf{A}}_{k-2:0}, \hat{\mathbf{A}}'_{k-2:0}, k \in \{1,\dots, N\}$, where the proportionality appears instead of an equality, as the operations $\hat{\mathcal{C}}^{(ij)}_{k-1}$ are not necessarily trace preserving. If Eq.~\eqref{eqn::Markov} is not satisfied, then the process is non-Markovian, since the only way past actions could influence the future evolution is through some kind of memory. Here, the sequence of operations $\mathbf{A}_{k-2:0}$ until $t_{k-2}$ together with the measurement outcome $\Pi^{(j)}_{k-1}$ constitute the `history' or `trajectory' of the system, whereas $P_{k-1}^{(i)}$ is its state at $t_{k-1}$. The concept of operationally well-defined quantum trajectories has recently been used to define Markovianity in an equivalent way to Eq.~\eqref{eqn::Markov}~\cite{sakuldee_non-markovian_2018}. Since $\hat{\mathcal{T}}^{k:0}$ is a linear operator, Eq.~\eqref{eqn::Markov} allows the process to be checked for Markovianity with a finite number of experiments~\cite{pollock_operational_prl, pollock_complete_pra}. As for the case of initial correlations, the unambiguous detection of memory effects necessitates full experimental control over the system of interest. 

For the investigation of the Markovianity of a process by means of $\mathcal{T}^{N:0}_\mathcal{F}$, there are -- just like for the case of correlations -- two cases that have to be distinguished. Whenever $\text{Span}\left(\mathcal{F}\right)$ contains at least one causal break, Eq.~\eqref{eqn::Markov} can be evaluated directly. If there are at least two different `histories', such that the final state on the system does not only depend on the freshly prepared state, the process is non-Markovian. As before, this can be tested for in a finite number of experiments, and as for the case of correlations, the converse does not hold; if the restricted set of operations fails to detect a history dependence/memory, it does not mean that there is none. An example of an experimental setup for which $\text{Span}\left(\mathcal{F}\right)$ contains causal breaks, but does not allow to unambiguously decide for the existence of memory effects, is one in which the experimenter can only perform projective measurements (see Sec.~\ref{subsec::Proj}). 

The second case that has to be investigated is the one where $\text{Span}\left(\mathcal{F}\right)$ does not contain causal breaks and a direct evaluation of Eq.~\eqref{eqn::Markov} is impossible. A prominent example of this is the case where $\mathcal{F}$ is the set of unitary operations (see Sec.~\ref{sec:unitaries}). As the detection of non-Markovianity hinges on the availability of causal breaks, such a set of local operations seems to be inadequate for the investigation of the Markovianity of a process. However, as discussed above, it can be used for the detection of system-environment correlations.
Intuitively, it is clear that, if correlations can be detected at any time via local operations, the process must be non-Markovian. Correlations constitute a memory of past interactions, and their detection implies that memory plays a non-negligible role in the process. We prove this statement formally in App.~\ref{App::ProofMark}. Consequently, non-Markovianity can be tested for by checking for system-environment correlations at each time $t_j$, for varying earlier sequences of operations $\mathbf{A}_{j-1:0} \in \left[\mathrm{Span}(\mathcal{F})\right]^{\otimes j}$. Again, the converse statement does not hold. On the one hand, a restricted set of local operations does not necessarily allow for the detection of all locally detectable correlations. On the other hand, a process can also be non-Markovian without any system-environment correlations being present at any point in time~\cite{pollock_operational_prl}.

\section{Examples and Applications}
\label{sec::Examples}
\subsection{Reconstruction of qubit dynamics}
\label{sec:XTomo}
It is instructive to illustrate the reconstruction of restricted process tensors and correlation witnesses with a low dimensional and computationally accessible example. While the number of necessary measurements for the reconstruction of a process tensor depends on the size of the system, we emphasize that independent of the size of the system or the environment, and the type of interaction between them our framework can in principle always be experimentally reconstructed. We merely choose low-dimensional examples to keep analytical calculations compact and simple. For ease of notation, for the most part, we restrict the investigation to the one-step process tensors $\mathcal{T}^{1:0}_{\mathcal{U}}$ and $\mathcal{T}^{1:0}_\mathcal{P}$ constructed from local unitary maps and projective maps, respectively. The generalization to restricted $N$-step process tensors is always possible in a straight forward way. 

The simplest conceivable open quantum system is the case of a 
qubit coupled to another qubit that serves as the environment. While in general, the initial system-environment state before preparation is considered constant and part of the dynamics (see Fig.~\ref{fig::ProcTens}), here, we investigate a family of initial states in order to explicitly analyze the set of detectable correlations. We choose the seven-dimensional family of X-states~\cite{yu_evolution_2007,ali_quantum_2010} as a blueprint for initial states $\rho_{\mathcal{SE}}$. X-states are states of the form 
 \begin{gather}
  \label{eqn::Xstates}
  \rho^X_{\mathcal{SE}} = \begin{pmatrix} a_{11}&0&0&a_{14} \\ 0&a_{22}&a_{23}&0\\ 0&a_{23}^*&a_{33}&0 \\ a_{14}^*&0&0&a_{44} \end{pmatrix}
 \end{gather}
in a given basis (in our case the eigenbasis of $\sigma_z \otimes \sigma_z$). Eq.~\eqref{eqn::Xstates} describes a valid quantum state iff $\rho_X^{\mathcal{SE}}$ has unit trace, $a_{22}a_{33} \geq 
\left|a_{23}\right|^2$
 and $a_{11}a_{44} \geq \left|a_{14}\right|^2$.
 The family of X-states includes both entangled states and separable states, as well as product states. Hence, it is well-suited for the analysis of the detectability of correlations by means of local unitary operations or projective measurements. For our system-environment Hamiltonian we choose
 \begin{gather}
  \label{eqn::Hamiltonian}
  H_{\mathcal{SE}} = \omega\left(\sigma_x \otimes \sigma_x + \sigma_y \otimes \sigma_y + \sigma_z \otimes \sigma_z\right),
 \end{gather}
where $\omega \in \mathbb{R} \setminus\left\{0\right\}$. These choices of $H_{\mathcal{SE}}$ and $\rho^X_{\mathcal{SE}}$ allow for an analytical construction of $\mathcal{T}^{1:0}$, $\mathcal{T}^{1:0}_{\mathcal{U}}$ and $\mathcal{T}^{1:0}_{\mathcal{P}}$, but are also non-trivial enough to exhibit the features we would like to investigate, in particular the effects of initial system-environment correlations. 

With this, we can explicitly construct the full one-step process tensor, as well as the restricted ones. To this end, we compute the outcome states for a full basis of the space $\mathcal{B}\left(\mathcal{H}_\mathcal{S}\right)\otimes\mathcal{B}\left(\mathcal{H}_\mathcal{S}\right)$ of local operations, where, in this case, $\mathcal{H}_{\mathcal{S}} \cong \mathbbm{C}^2$. Let $\left\{\sigma_x,\sigma_y,\sigma_z\right\}$ be the Pauli matrices. A possible basis of $\mathcal{B}\left(\mathcal{H}_\mathcal{S}\right)\otimes\mathcal{B}\left(\mathcal{H}_\mathcal{S}\right)$ is given by the set $\left\{Q_i \otimes Q_j^\text{T}\right\}$, where the pure states
\begin{align}
\notag
 &Q_1 = \frac{1}{2} \left(\mathbbm{1} + \sigma_z\right), \quad 
 Q_2 = \frac{1}{2} \left(\mathbbm{1} + \sigma_x\right), \\ \quad &Q_3 = \frac{1}{2} \left(\mathbbm{1} - \sigma_x\right), \quad   Q_4 = \frac{1}{2} \left(\mathbbm{1} + \sigma_y\right)\, ,
\end{align}
 constitute a basis of $\mathcal{B}\left(\mathcal{H}_\mathcal{S}\right)$. The corresponding output states $\eta'_{(jk)}$ at time $t$ are given by~\footnote{Here and in what follows we put brackets around the subscripts to clearly distinguish them from matrix elements whenever necessary.}
\begin{gather}
 \eta'_{(jk)} = \tr_\mathcal{E} \left\{e^{-\text{i}H_{\mathcal{SE}}t}\left(\hat{\mathcal{Q}}_{(jk)} \otimes \mathcal{I}_{\mathcal{E}}\right)\left[\rho^X_{\mathcal{SE}}\right]e^{\text{i}H_{\mathcal{SE}}t}\right\},
\end{gather}
where the action of $\hat{\mathcal{Q}}_{(jk)}$ on a state $\rho \in \mathcal{B}(\mathcal{H}_\mathcal{S})$ is given by $\hat{\mathcal{Q}}_{(jk)}[\rho] = \tr(Q_k\rho)Q_j$. The full one-step process tensor $\mathcal{T}^{1:0}$ can be calculated using Eq.~\eqref{eqn::ProcTensRec} as 
\begin{gather}
 \mathcal{T}^{1:0} = \sum_{j,k=1}^4 \eta'_{(jk)} \otimes \Delta_j^\text{T} \otimes \Delta_k^\text{T}\, ,
\end{gather}
with $\tr\left( \Delta_l Q_m\right) = \delta_{lm}\,$. 
Accordingly, in order to construct $\mathcal{T}^{1:0}_{\mathcal{U}}$ based on the set of unitary preparations, we fix a basis of the space spanned by the unitary maps. Here, we choose the basis $\left\{\mathcal{Z}_{\alpha}\right\}=\left\{\mathcal{Z}_{0} = \mathcal{I},\mathcal{Z}_{(j,\pm)},\mathcal{Z}_{(j+k+1,+)} \right\}$ (with $j,k \in \left\{1,2,3\right\}$ and $j<k$), derived in App.~\ref{app::BasisUnitQubit}. However, in principle, any set of $(d_\mathcal{S}^2-1)^2 + 1 = 10$ linearly independent unitary operations would yield the same results. 
The output states corresponding to each of these unitary preparations are given by
\begin{gather}
 \zeta'_{\alpha} =  \tr_\mathcal{E} \left\{e^{-\text{i}H_{\mathcal{SE}}t}\left(\hat{\mathcal{Z}}_{\alpha} \otimes \mathcal{I}_{\mathcal{E}}\right) \left[\rho^X_{\mathcal{SE}}\right]e^{\text{i}H_{\mathcal{SE}}t}\right\}\, ,
\end{gather}
The restricted process tensor $\mathcal{T}_{\mathcal{U}}^{1:0}$ can be constructed via
\begin{gather}
 \mathcal{T}_{\mathcal{U}}^{1:0} = \sum_{\alpha=1}^{10} \zeta'_\alpha \otimes \mathcal{X}_\alpha^\text{T}, \quad \text{where} \quad \tr\left( \mathcal{X}_\alpha\mathcal{Z}_{\beta}\right) = \delta_{\alpha\beta}
\end{gather}
Analogously, we construct $\mathcal{T}^{1:0}_\mathcal{P}$ by using the basis of the set of projective maps introduced in Eq.~\eqref{eqn::PureStates}. Given $\mathcal{T}^{1:0}$, the correlation-memory matrix $\mathcal{K}$ can be readily derived; the restricted correlation-memory matrix $\mathcal{K}_{\mathcal{U}}$ ($\mathcal{K}_\mathcal{P}$) is obtained by constructing the one-step process tensor $\mathcal{L}^{1:0}_\mathcal{U}$ ($\mathcal{L}^{1:0}_\mathcal{P}$) for the uncorrelated initial state $\rho^X_{\text{Prod.}} = \rho^X_\mathcal{S} \otimes \rho^X_{\mathcal{E}}$ and subtracting it from $\mathcal{T}_\mathcal{U}^{1:0}$ ($\mathcal{T}^{1:0}_\mathcal{P})$. 

It turns out that, except for trivial total dynamics ($\omega t = n\frac{\pi}{2}$), the restricted correlation-memory matrices $\mathcal{K}_{\mathcal{U}}$ and $\mathcal{K}_{\mathcal{P}}$ are equal to zero iff 
\begin{gather}
 \label{eqn::CondZero}
 a_{23} = a_{14} = 0 \quad \text{and} \quad a_{22}a_{33} = a_{11}a_{44}. 
\end{gather}
We have $\mathcal{K}_{\mathcal{V}} \neq \mathcal{K}_{\mathcal{P}}$ and the total correlation-memory matrix $\mathcal{K}$ differs both from $\mathcal{K}_{\mathcal{V}}$ and $\mathcal{K}_{\mathcal{P}}$; however, $\mathcal{K}$ is also equal to zero iff Eq.~\eqref{eqn::CondZero} is fulfilled.
X-states that satisfy the relations in Eq.~\eqref{eqn::CondZero} are product states which means that in this particular case, any correlations (classical or quantum) between the system and the environment could be detected with both $\mathcal{K}_\mathcal{U}$ and $\mathcal{K}_{\mathcal{P}}$ (and hence with the unrestricted correlation-memory matrix $\mathcal{K}$). Note that the `do nothing' correlation-memory matrix $\mathcal{K}_{\mathcal{I}_\mathcal{S}}$, \textit{i.e.}, the correlation-memory matrix that can be constructed without performing any local operation, would be zero iff $\mathfrak{Im}\left(a_{23}\right)=0$ (\textit{i.e.}, $a_{23} \in \mathbb{R}$). Performing local operations substantially increases the set of detectable correlations. 

In this example, unitary preparation and projections can reveal exactly the same kinds of initial correlations as the full set of local operations. This is not generally true. Consider, for example, a two-qubit state with correlation matrix $\chi_{\mathcal{SE}} = \alpha\sigma_z \otimes \sigma_x$, where $\alpha \in \mathbb{R}\setminus \left\{0\right\}$, and a total unitary evolution that is given by the swap operator $S$. The action of $\mathcal{K}_\mathcal{U}$ on an arbitrary unitary map $\mathcal{\mathcal{Z}}$ is then given by 
\begin{align}
\notag
 \hat{\mathcal{K}}_{\mathcal{U}}[\hat{\mathcal{Z}}] &= \alpha \tr_{\mathcal{E}}\left\{S\left(\hat{\mathcal{Z}}\left[\sigma_z\right] \otimes \sigma_x\right)S^{\dagger}\right\} \\ &= \alpha \tr_{\mathcal{E}}\left\{ \sigma_x\otimes \hat{\mathcal{Z}}\left[\sigma_z\right]\right\} = 0,
\end{align}
and hence $\mathcal{K}_\mathcal{U} = \mathbf{0}$. On the other hand, $\mathcal{K} \neq \mathbf{0}$; for example, if we consider the causal break $\mathcal{C} = \ket{0}\bra{0} \otimes \ket{1}\bra{1}$, we obtain
\begin{align}
\notag
 \hat{\mathcal{K}}[\hat{\mathcal{C}}] &= \alpha\tr_{\mathcal{E}}\left\{S\left[\ket{0}\bra{0} \bra{1}\sigma_z \ket{1} \otimes \sigma_x\right]S^\dagger\right\} \\
 &= -2\alpha\tr_{\mathcal{E}}\left(\sigma_x \otimes Q_1\right) = -2\alpha\sigma_x.
\end{align}
In this case, the fact that the correlations cannot be detected by unitary preparations alone, stems from a particular interplay between the total unitary evolution (the swap operation $S$) and the correlation matrix $\chi_{\mathcal{SE}}$. 

\subsection{Restricted process tensors and quantum control}
An important field where a restricted set of performable local operations and the presence of non-Markovian effects come into play together is that of quantum control of open systems (see for example Ref.~\cite{altafini_modeling_2012} for an introduction). Here, generally speaking, the goal is to steer the system of interest to a desired final state by means of local, time-dependent Hamiltonians, which can be controlled by the experimenter. For microscopic models that assume knowledge of the total system-environment Hamiltonian, the impact of these local Hamiltonians on the dynamics of the system can often be readily deduced. However, if only local information, for example a master equation description, is at hand, it is in general unclear how to include the influence of a local operation into the description~\cite{arenz_distinguishing_2015}. The process tensor approach is tailored to solve this problem operationally. 

The presence of memory effects is of particular importance for dynamical decoupling experiments~\cite{viola_dynamical_1999}, where the local Hamiltonians are employed in such a way that they average out the influence of the environment and, effectively, decouple a system from its environment. This is only possible if memory effects are present~\cite{arenz_distinguishing_2015}.

Under the assumption that the time span over which the local Hamiltonians act is small compared to typical time scales of the dynamics of the system (\textit{i.e.} the Hamiltonians basically act at fixed times as `kicks' of infinite strength~\cite{viola_dynamical_1998}), such an experimental setup can be described as a mapping from a set of unitary maps $\left\{\hat{\mathcal{Z}}_{i}\right\}_{i=1}^{N-1}$, that act on the system at times $t_i$, to a final state $\rho'_{\mathcal{S}}\left(\mathcal{Z}_{1},\cdots,\mathcal{Z}_{N-1}\right)$. 

To demonstrate the applicability of the restricted process tensor framework to quantum control and dynamical decoupling, we show how it can be used to find an ideal decoupling sequence in the scenario where decoupling is required at a single fixed time $t_N$ (as opposed to decoupling for all times). We say that a sequence of unitary maps decouples the system from its environment at the time $t_N$ if the state of the system at $t_N$ is unitarily invariant to the input state, \textit{i.e.}, $\rho'\left(\mathcal{Z}_{1},\cdots,\mathcal{Z}_{N-1}\right) = \hat{\mathcal{Z}}\left[\rho_{\text{init.}}\right]$ for all initial system states $\rho_{\text{init.}}$, where $\hat{\mathcal{Z}}$ is a fixed unitary, and hence reversible, map.

In order for dynamical decoupling to be operationally meaningful, it has to be assumed that the initial total state is of product form (\textit{i.e.}, $\rho_{\mathcal{SE}} = \rho_{\text{init.}}\otimes \rho^\mathcal{E}_{\text{init.}}$), where the experimenter has control over the system state, and $\rho^\mathcal{E}_{\text{init.}}$ is independent of the choice of $\rho_{\text{init.}}$. The process tensor that describes the dynamics of the system is then a map from the controllable inputs (i.e. the initial system states and the sequence $\mathbf{Z}_{N-1:1}$ of unitary maps $\hat{\mathcal{Z}}_{i}$) to the final state $\rho'\left(\rho_{\text{init.}},\mathbf{Z}_{N-1:1}\right)$. Its experimental reconstruction follows from Eq.~\eqref{eqn::partProc} and is achieved by measuring the output states for a basis $\left\{\rho_\mu\right\}_{\mu=1}^{d_\mathcal{S}^2}$ of input states and a basis of sequences of unitaries, \textit{i.e.},
\begin{gather}
 \label{eqn::ProcTensDD}
 \mathcal{T}^{N:0}_\mathcal{U} = \sum_{\vec{i},\mu} \zeta'_{(\mu,\vec i\,)} \otimes \omega_\mu^\text{T} \otimes \mathcal{X}_{\vec i}^\text{T},
\end{gather}
where the maps $\left\{\omega_\mu\right\}_{\mu=1}^{d_\mathcal{S}^2}$  are the duals of $\left\{\rho_\mu\right\}_{\mu=1}^{d_\mathcal{S}^2}$, $\left\{\mathcal{X}_{\vec i}\right\}$ are the duals of a basis of sequences of unitary operations and $\zeta'_{(\mu,\vec i\,)}$ are the final system states for a choice $\left\{\rho_\mu, \mathcal{Z}_{i_1},\cdots,\mathcal{Z}_{i_N}\right\}$ of the initial system state and the $N-1$ local unitary operations. Given that the initial state is now an input of the process tensor, it is possible (see App.~\ref{app::procAction}) for any sequence of $N-1$ local unitary operations $\mathbf{Z}_{N-1:1} = \left\{\mathcal{Z}_{1},\cdots, \mathcal{Z}_{N-1}\right\}$ to construct the map $\hat{\mathcal{R}}_\mathbf{Z}: \mathcal{B}\left(\mathcal{H}_\mathcal{S}\right) \rightarrow \mathcal{B}\left(\mathcal{H}_\mathcal{S}\right)$, which depends on $\mathbf{Z}_{N-1:1}$ and maps initial states to final states:
\begin{gather}
 \label{eqn::dynamicMap}
 \hat{\mathcal{R}}_\mathbf{Z}\left[\cdot\right] = \hat{\mathcal{T}}^{N:0}_\mathcal{U} \left[\, \cdot\, ,\mathbf{Z}_{N-1:1}\right].
\end{gather}
This means that $\mathcal{R}_\mathbf{Z}$ is the resulting map acting on the initial state, given that the sequence $\mathbf{Z}_{N-1:1}$ of intermediary unitaries was performed~\footnote{In the language of Ref.~\cite{chiribella_quantum_2008}, we are contracting a quantum comb with another quantum comb to yield a channel.}. We therefore call a sequence $\mathbf{Z}_{N-1:1}$ decoupling if 
$\mathcal{R}_\mathbf{Z}$ is a unitary map. Given the restricted process tensor of a process, it is then merely a numerical sampling problem to find such a sequence (if it exists for the chosen time steps). To illustrate the description of dynamical decoupling in terms of a restricted process tensor, we reiterate the shallow pocket model discussed in Ref.~\cite{arenz_distinguishing_2015}.

Let $H_{\mathcal{SE}} = \frac{g}{2}\sigma_z \otimes \hat{x}$ be the total, time-independent Hamiltonian for a qubit coupled to a particle on a line. We choose $\rho \otimes \ket{\Psi}\bra{\Psi}$ as the initial states, where $\braket{x|\Psi} = \sqrt{\frac{\gamma}{\pi}} \frac{1}{x+\text{i}\gamma}$ ($\gamma >0$). The free evolution of the system state, \textit{i.e.} the evolution without intermediate local operations, is given by~\cite{arenz_distinguishing_2015}
\begin{gather}
 \label{eqn::freeEvo}
 \rho(t) = \left(\begin{array}{ll}
                        \rho_{00}(0) & \rho_{01}(0)e^{-g\gamma t} \\ \rho_{01}^*(0) e^{-g\gamma t} & \rho_{11}(0)
                       \end{array}\right),
\end{gather}
which constitutes a purely dephasing dynamics. A possible experimental decoupling procedure could consist of a choice of the initial system state $\rho_\mathcal{S}$, a free evolution of the system-environment state according to $ H_{\mathcal{SE}}$ for a time $\Delta t$, a local unitary operation $\mathcal{Z}$, and, finally, a tomography of the system state after another free evolution for a time $\Delta t$. Let $\mathcal{T}^{2:0}_\mathcal{U}$ be the restricted process tensor for this experiment. We have 
\begin{gather}
 \hat{\mathcal{T}}^{2:0}_\mathcal{U}\left[\rho_\mathcal{S},\hat{\mathcal{Z}}\right] = \tr_\mathcal{E}\left\{\hat{\mathcal{U}}  \left[\hat{\mathcal{Z}}\left[\, \hat{\mathcal{U}}\left[\rho_\mathcal{S} \otimes \ket{\Psi}\bra{\Psi}\right]\right]\right] \right\}, 
\end{gather}
where $\hat{\mathcal{U}}\left[\rho_{\mathcal{SE}}\right] = e^{-\text{i}H_\mathcal{SE}\Delta t} \rho_\mathcal{SE} e^{\text{i}H_\mathcal{SE}\Delta t}$ and we have omitted an identity map on the environment. Setting $Z_{(a,b)} = a\sigma_x + b\sigma_y$ and $\hat{\mathcal{Z}}_{(a,b)}\left[\rho\right] =Z_{(a,b)}\rho Z_{(a,b)}^\dagger$, where $\left|a\right|^2 + \left|b\right|^2 = 1$ and $a,b \in \mathbb{R}$, we obtain
\begin{gather}
 \label{eqn::ContractProcTens}
 \hat{\mathcal{T}}^{2:0}_\mathcal{U}\left[\, \cdot\, , \hat{\mathcal{Z}}_{(a,b)}\right] = \hat{\mathcal{Z}}_{(a,b)}^*\left[\, \cdot \,\right],
\end{gather}
with $\hat{\mathcal{Z}}_{(a,b)}^*\left[\rho_\mathcal{S}\right] = Z_{(a,b)}^\dagger\rho_\mathcal{S} Z_{(a,b)}$. Consequently, any local operation of the form $\hat{\mathcal{Z}}_{(a,b)}$ decouples the system of interest from its environment for the given process. This can also be shown directly from the total Hamiltonian $H_\mathcal{SE}$~\cite{arenz_distinguishing_2015}. However, the restricted process tensor can be reconstructed based on local operations alone and allows for a numerical search of a sequence $\mathbf{Z}_{N-1:1}$ of unitary operations, such that $\hat{\mathcal{R}}_\mathbf{Z}$ is a unitary map. 

On the other hand, a description of the dynamics in terms of a master equation would fail to reproduce these results (as shown in Ref.~\cite{arenz_distinguishing_2015}). The master equation of the free open evolution of the system is given by 
\begin{gather}
 \label{eqn::Lindbladian}
 \dot\rho=\hat{L}\left[\rho\right] = -g\frac{\gamma}{4}\left[\sigma_z,\left[\sigma_z,\rho\right]\right]
\end{gather}
$\forall \rho \in \mathcal{B}\left(\mathcal{H}_\mathcal{S}\right),$ where $\hat{L}$ is the Lindbladian of the time evolution. The dynamics in Eq.~\eqref{eqn::freeEvo} is then given by $\rho\left(t\right) = e^{\hat{L}t}\left[\rho(0)\right]$, and, naively, the dynamics including the intermediate local operation $\hat{\mathcal{Z}}_{(a,b)}$ would be described by $\rho_{(L)}\left(2\Delta t\right) =\left(e^{\hat{L}\Delta t} \hat{\mathcal{Z}}_{(a,b)}e^{\hat{L}\Delta t}\right)\left[\rho(0)\right]$. Given that $e^{\hat{L}\Delta t}$ is a purely dephasing channel, we conclude that $\rho_{(L)}\left(2\Delta t\right)$ is generally not equal up to unitary transformation to the initial system state $\rho(0)$. As soon as the dynamics are non-Markovian, standard master equations fail to capture the influence that local operations at intermediate time steps have on the dynamics of the system of interest.

\section{Conclusion and Outlook}
General non-Markovian quantum dynamics can be unambiguously described and characterized experimentally if the experimenter has unlimited control, \textit{i.e.}, access to a basis of the space of possible operations on the system of interest. In this paper, we have investigated the more realistic situation, where the set of accessible operations is restricted by the experimental setup. We have shown that in these cases, it is still possible -- independent of the existence of memory effects -- to reconstruct a process tensor, as long as the dimension of the space that is spanned by the available operations is known. The obtained restricted process tensor contains the maximal amount of information about the process, that can be inferred locally, based on the set of available operations. Unlike for full process tensors, complete positivity is in general not a well-defined property for restricted process tensors, yet they still satisfy the containment property. Restricted process tensors can be applied to any operation that lies in $\text{Span}(\mathcal{F})$ and provides an operationally meaningful complete dynamical description of the underlying dynamics. Surprisingly, the set of operations a restricted process tensor can be applied to can exceed the experimentally available ones. For example, if the set of available manipulations coincides with the set of unitary maps, the reconstructed process-tensor can, \textit{e.g.}, be applied to any sequence of unital operations.

We demonstrated that if one further local operation, a swap with an identically prepared system, is performable, or if the span of the set of performable operations contains operations that decouple the system from its environment, it is possible to construct operationally well-defined witnesses for initial system-environment correlations and the non-Markovianity of a process. We have shown how to make maximal use of the available operations in the construction of these witnesses and illustrated their applicability for two extremal cases: the set of unitary operations, where no information about the system can be inferred from the operation, and the set of projective measurements, where information about the system is obtained, but it collapses to a pure state in the process. In both cases, the reconstructed witnesses detect initial correlations, as well as the non-Markovianity of the underlying process.

The quality of these witnesses, \textit{i.e.}, their ability to detect correlations, depends crucially on the dimension of the space that is spanned by the available local operations and the interplay between correlations and the total unitary dynamics. However, we conjecture that for any reasonable scenario, \textit{i.e.} a general total unitary dynamics and a set of performable local operations that is not `too small', it is always possible to detect system-environment correlations by means of the experimentally realizable local operations. Total unitary dynamics that prevent correlations from local detection should be mere pathological examples. We defer a thorough numerical investigation of this statement to future works. 

Restricted process tensors bridge the gap between the theoretical description of non-Markovian quantum dynamics and experimental reality. Furthermore, they provide the ideal framework for the description of quantum control and dynamic decoupling. In order to simplify the calculations, we have demonstrated this for a time-independent total Hamiltonian. In practice, however, the restricted process tensor can be reconstructed experimentally for any conceivable total dynamics. Given the restricted process tensor, it is then simply a numerical sampling problem to find the optimal sequence of operations that steers the system as close as possible to to a desired final state. Our framework is also versatile enough to describe decoupling experiments and it allows one to search for the sequence of local operations that comes closest to achieving decoupling at a fixed time $t_N$. While this is not the original aim of dynamical decoupling, it nonetheless provides a new perspective: if decoupling at selective points in time is sufficient, decoupling schemes based on restricted process tensors might prove more efficient and less error-prone than traditional schemes that rely on the implementation of decoupling cycles much faster than typical correlation times~\cite{viola_dynamical_1999}. 

Even if perfect decoupling will in general not be possible for randomly chosen time steps $t_k$, the process tensor approach is nonetheless fruitful: it opens up an avenue to benchmarking the deviation from perfect decoupling for a given choice of time steps and translates the question of whether perfect decoupling is possible to an inversion problem of the process tensor for the underlying process.

%\begin{acknowledgements}
{\bf Acknowledgements.---}
We are grateful to F. Sakuldee for valuable conversations. SM is supported by the Monash Graduate Scholarship (MGS), Monash International Postgraduate Research Scholarship (MIPRS) and the J L William Scholarship.
%\end{acknowledgements}

\appendix
\onecolumngrid
\section{Action of a process tensor on \texorpdfstring{$\hat{\mathbf{A}}_{N-1:0}$}{}}
\label{app::procAction}
Let $\mathcal{T}^{1:0} \in \mathcal{B}(\mathcal{H}_\text{out})\otimes \mathcal{B}(\mathcal{H}^0_\text{out})\otimes \mathcal{B}(\mathcal{H}^0_\text{in}) $ be the Choi state of a one-step process tensor. Its action on a completely positive map $\hat{\mathcal{A}}$ is given by~\cite{modi_operational_2012}:
\begin{gather}
 \label{eqn::SuperAction}
 \hat{\mathcal{T}}^{1:0}\left[\hat{\mathcal{A}}\right] = \tr_{\text{in}}\left[(\mathbbm{1}_\text{out} \otimes \mathcal{A}^{\mathrm{T}})\mathcal{T}^{1:0}\right]\, ,
\end{gather}
where $\mathcal{A} \in \mathcal{B}(\mathcal{H}^0_{\text{out}})\otimes \mathcal{B}(\mathcal{H}^0_{\text{in}})$ and $\tr_{\text{in}}$ denotes the trace over the Hilbert space $\mathcal{H}^0_{\text{out}}\otimes \mathcal{H}^0_{\text{in}}$. Here, for clarity, we enumerate the respective Hilbert spaces by the time-step they belong to, \textit{i.e.}, $\mathcal{B}(\mathcal{H}^k_{\text{out}})\otimes \mathcal{B}(\mathcal{H}^k_{\text{in}})$ is the space that contains the Choi states of CP operations at time $t_k$, while the final output state lies in $\mathcal{B}(\mathcal{H}_\text{out})$. Eq.~\eqref{eqn::SuperAction} generalizes to the multi-time step case in a straight forward fashion. Using Eq.~\eqref{eqn::SuperAction}, it can be readily shown that Eq.~\eqref{eqn::ProcTensRec} yields the correct process tensor of a process. We show this explicitly for the one-step case: The Choi state of any completely positive map $\hat{\mathcal{A}}$ can be written as $\mathcal{A} = \sum_{\mu=1} a_\mu D_\mu$, where $a_\mu \in \mathbb{R}$ and $\left\{D_\mu\right\}_{\mu=1}^{d_\mathcal{S}^4}$ is a basis of $\mathcal{B}(\mathcal{H}^0_{\text{out}})\otimes \mathcal{B}(\mathcal{H}^0_{\text{in}})$. Let $\left\{\Delta_\nu\right\}_{\nu=1}^{d_\mathcal{S}^4}$ be the dual set to this basis. Using Eq.~\eqref{eqn::SuperAction}, the action of $\mathcal{T}^{1:0}$ constructed according to Eq.~\eqref{eqn::ProcTensRec} on $\mathcal{A}$ is then given by:
\begin{align}
 \hat{\mathcal{T}}^{1:0}\left[\hat{\mathcal{A}}\right] = \sum_{\mu,\nu=1} a_\mu \tr_{\text{in}}\left[(\mathbbm{1}_\text{out} \otimes D_\mu) \rho'_{\nu} \otimes \Delta_\nu\right] = \sum_{\mu,\nu=1} a_\mu \rho'_\mu \tr(D_\mu  \Delta_\nu) = \sum_{\mu=1} a_\mu \rho'_\mu = \rho'(\mathcal{A})
\end{align}
Any process tensor can be contracted with a set of local operations to yield another process tensor. For example, let $\mathcal{T}^{2:0}$ be a two-step process tensor. Contraction with a local operation  $\mathcal{A}_0$ (performed at $t_0$) yields 
\begin{gather}
 \label{eqn::Contraction}
 \mathcal{T}^{2:1|\mathcal{A}_{0}} = \tr_0\left[(\mathbbm{1}_\mathrm{out} \otimes \mathbbm{1}^1 \otimes \mathcal{A}_0^{\mathrm{T}})\mathcal{T}^{2:0}\right]\, ,
\end{gather}
where $\mathbbm{1}^1\in \mathcal{B}(\mathcal{H}^1_{\text{out}})\otimes \mathcal{B}(\mathcal{H}^1_{\text{in}})$ is an identity matrix and $\tr_0$ denotes the trace over the Hilbert space $\mathcal{H}^0_{\text{out}}\otimes \mathcal{H}^0_{\text{in}}$. The contracted process tensor $\mathcal{T}^{2:1|\mathcal{A}_0}$ describes the dynamics of the underlying process, given that the local operation $\mathcal{A}_0$ was performed in the first time $t_0$. Consequently, $\mathcal{T}^{2:1|\mathcal{A}_0}$ can be applied to any local operation $\mathcal{A}_1$ that can be performed at the $t_1$ time step:
\begin{gather}
 \hat{\mathcal{T}}^{2:1|\mathcal{A}_0}\left[\hat{\mathcal{A}}_1\right] = \rho'\left(\mathcal{A}_0,\mathcal{A}_1\right)\, ,
\end{gather}
and yields the correct output state. For a more detailed discussion of the mathematical properties of quantum maps and process tensors, see Ref.~\cite{milz_introduction_2017} and the appendices of Ref.~\cite{pollock_operational_prl}.

\section{Derivation of intermediate restricted process tensors}
\label{app::ICPOVM}
Here, we show how to obtain intermediate restricted process tensors in the case where $\text{Span}(\mathcal{F})$ contains an informationally complete POVM. We emphasize that the following derivation does \textit{not} apply if the available operations do not allow for a full tomography of the state of the system at each time step. 

Let the action of the set $\{\mathcal{A}_\mu\} \subset \text{Span}(\mathcal{F})$ be an IC POVM on the system, \text{i.e.}, $\hat{\mathcal{A}}_\mu[\rho] = E_\mu\rho E_\mu^\dagger$, where the corresponding positive operators $\{\Pi_\mu=E_\mu^\dagger E_\mu\}$ satisfy $\sum_\mu \Pi_\mu = \mathbbm{1}_\mathcal{S}$ and every state $\rho$ is uniquely determined by the probabilities $p_\mu = \tr\{\hat{\mathcal{A}}_\mu[\rho]\}=\tr(\Pi_\mu \rho)$. In order to derive $\mathcal{T}^{k:0}$ ($0<k<N$) from $\mathcal{T}^{N:0}$, the output states at $t_k$ corresponding to a basis of input sequences $\mathbf{A}_{k-1:0}$ have to be determined. 
Let $\mathbf{A}'_{k-1:0} \in \text{Span}(\mathcal{F})^{\otimes k}$ be a fixed sequence of operations. We obtain
\begin{gather}
 \hat{\mathcal{T}}_\mathcal{F}^{N:0}\left[\mathcal{I}_\mathcal{S}^{\otimes (N-k-1)}, \hat{\mathcal{A}}_\mu, \hat{\mathbf{A}}'_{k-1:0}\right] = p_\mu p' \eta_\mu \, ,
\end{gather}
where $\eta_\mu$ is a unit trace quantum state that depends on $\hat{\mathcal{A}}_\mu$, $\hat{\mathbf{A}}'_{k-1:0}$ and $\mathcal{I}_\mathcal{S}^{\otimes (N-k-1)}$; $p'$ is the trace of the system's state $\rho' := p' \eta'$ at $t_k$ after the sequence of operations $\mathbf{A}'_{k-1:0}$ was performed and $p_{\mu} = \tr\{\hat{\mathcal{A}}_\mu [\eta']\}$. For fixed  $\mathbf{A}'_{k-1:0}$, we have $\sum_\mu \tr(p_{\mu} p' \eta_\mu) = p'$, as $\sum_\mu p_\mu=1$. Consequently, all probabilities $p_\mu$ can be derived via
\begin{gather}
 p_\mu = \frac{1}{p'}\tr\left\{\hat{\mathcal{T}}_\mathcal{F}^{N:0}\left[\mathcal{I}_\mathcal{S}^{\otimes (N-k-1)}, \hat{\mathcal{A}}^{(\mu)}, \hat{\mathbf{A}}'_{k-1:0}\right]\right\}\, ,
\end{gather}
which unambiguously determines the state $\rho'$ at the time step $t_k$ after the sequence of operations $\hat{\mathbf{A}}'_{k-1:0}$ was performed. Accordingly, we can deduce the output states at $t_k$  corresponding to a basis of input sequences $\mathbf{A}_{k-1:0}$, which enables the construction of $\mathcal{T}^{k:0}_\mathcal{F}$. Note that the above construction also works if the sequence $\mathcal{I}_\mathcal{S}^{\otimes (N-k-1)}$ of `do-nothing' operations at the time steps $\{t_{k+1},\dots,t_{N-1}\}$ is replaced by a fixed sequence of trace preserving CP maps. From $\hat{\mathcal{T}}_\mathcal{F}^{k:0}$, we obtain $\hat{\mathcal{T}}_\mathcal{F}^{k:j}[\cdot] = \hat{\mathcal{T}}_\mathcal{F}^{k:0}[\, \cdot \, , \mathcal{I}_\mathcal{S}^{\otimes j}]$, where $0 \leq j < k$. 
\section{Basis of unitary maps acting on a qubit}
\label{app::BasisUnitQubit}
Any unitary matrix $Z \in \text{SU(2)}$ can be expressed in terms of Pauli matrices $\left\{\sigma_i\right\}$ (where $i=x,y,z$) in the following form:
\begin{gather}
 \label{eqn::unitary_mat}
 Z_{(\alpha,\vec a)}= \cos\left(\frac{\alpha}{2}\right)\mathbbm{1} - \mathrm{i} \sin\left(\frac{\alpha}{2}\right) \sum_{i=1}^3 a_i \sigma_i, 
\end{gather}
where $\left|\vec a\right| = 1$. Hence, a generic unitary $1$-qubit map is of the form
\begin{align}
 \hat{\mathcal{Z}}_{(\alpha,\vec a)}\left[\rho\right] =& \cos^2\left(\frac{\alpha}{2}\right) \underbrace{\mathbbm{1} \rho \, \mathbbm{1}}_{(I)}
 + \sin^2\left(\frac{\alpha}{2}\right) \sum_{i=1}^3 a_i^2 \underbrace{\left( \sigma_i \rho \, 
 \sigma_i \right)}_{(II)} +   \cos\left(\frac{\alpha}{2}\right)\sin\left(\frac{\alpha}{2}\right)  \sum_{i=1}^{3}a_i \underbrace{\text{i}\left(
 \mathbbm{1} \rho \, \sigma_i - \sigma_i \rho \, \mathbbm{1} \right)}_{(III)}   \notag \\
& + \sin^2\left(\frac{\alpha}{2}\right) 
\sum_{i<k}^3 a_i a_k \underbrace{\left( \sigma_i \rho \, \sigma_k + \sigma_k \rho \, \sigma_i\right)}_{(IV)}.
\label{eqn::unitMap}
\end{align}
The term (I) in Eq.~\eqref{eqn::unitMap} can be accounted for by choosing $Z_{0} = \mathbbm{1}$. Following the analogous derivation for the case of projective maps in~\cite{kuah_how_2007}, we set 
\begin{gather}
\label{eqn::unitBasis1} 
 Z_{(j,\pm)} = \frac{1}{\sqrt{2}}\left(\mathbbm{1} \pm \mathrm{i}\sigma_j\right).
\end{gather}
With the six matrices given in Eq.~\eqref{eqn::unitBasis1} both (II) and (III) can be obtained:
\begin{gather}
 \sigma_j \rho \sigma_j = 2\left(\hat{\mathcal{Z}}_{(j,+)} + \hat{\mathcal{Z}}_{(j,-)}\right)\left[\rho\right] - \hat{\mathcal{Z}}_{0}\left[\rho\right]  \quad \text{and} \quad
  \text{i}\left(\mathbbm{1} \rho \, \sigma_j - \sigma_j \rho \mathbbm{1}\right) = 2\left(\hat{\mathcal{Z}}_{(j,-)}- \hat{\mathcal{Z}}_{(j,+)}\right)\left[\rho\right].
\end{gather}
The three remaining terms (IV) can be obtained with the three additional unitary matrices
\begin{gather}
 \label{eqn::unitBasis2}
 Z_{(j+k+1,+)} = \frac{1}{\sqrt{2}}\left(\mathbbm{1} + \frac{\text{i}}{\sqrt{2}}\sigma_j + \frac{\text{i}}{\sqrt{2}}\sigma_k \right) \, , 
\end{gather}
with $j<k$. We have
\begin{align}
\notag
 \sigma_j \rho \, \sigma_k + \sigma_k \rho \, \sigma_j = 2\left\{2\,\hat{\mathcal{Z}}_{(j+k+1,+)} 
 - \left(1+\sqrt{2}\right)\left(\hat{\mathcal{Z}}_{(j,+)}
 %-  \left(1+\sqrt{2}\right)
 +\hat{\mathcal{Z}}_{(k,+)}\right)
 - \left(1-\sqrt{2}\right)\left(\hat{\mathcal{Z}}_{(j,-)} 
 %- \left(1-\sqrt{2}\right)
 +\hat{\mathcal{Z}}_{(k,-)}\right)
 \right\} \left[\rho\right].
\end{align}
Hence, any unitary map $\hat{\mathcal{Z}}\left[\rho\right] = Z\rho \, Z^{\dagger}$ acting on a qubit can be represented as a linear combination of the ten unitary maps $\left\{\hat{\mathcal{Z}}_{0},\hat{\mathcal{Z}}_{(j,\pm)},\hat{\mathcal{Z}}_{(j+k+1,+)}\right\}$. A one-step process tensor constructed based on this set of operations can meaningfully be applied to any completely positive map that lies in its linear span, which, in this case, is the set of all one-qubit unital maps.

\section{Set of projective maps}
\label{app::Proj}

 A rank-$1$ projective map $\hat{\mathcal{Q}}\in \mathcal{P}$ acting on a $d_\mathcal{S}$-dimensional state $\rho_\mathcal{S}$ can be written as:
\begin{gather}
 \hat{\mathcal{Q}}\left[\rho_\mathcal{S}\right] = Q\rho_\mathcal{S}Q,
\end{gather}
where $Q = \sum_{k,l=1}^{d_\mathcal{S}} c_k c_l^* \ket{k}\bra{l}$ is a ($d_\mathcal{S}$-dimensional) pure state and ${}^*$ denotes complex conjugation. The Choi matrix of the map $\hat{\mathcal{Q}}$ has the form $Q \otimes Q^{\text{T}}$. Any map $\hat{\mathcal{N}} \in \text{Span}(\mathcal{P})$ can be represented as
\begin{gather}
\label{eqn::vecProj}
 \mathcal{N} = \sum_\nu b_\nu Q_{\nu} \otimes Q_{\nu}^\text{T} = \sum_{\nu}b_\nu\sum_{\substack{k,l=1 \\k',l'=1}}^{d_\mathcal{S}} c_k^{(\nu)} c_l^{(\nu)^*} c_{k'}^{(\nu)^*}c_{l'}^{(\nu)} \ket{k k'}\bra{ll'} \, .
\end{gather}
 The elements of any operator $\mathcal{N} \in \text{Span}(\mathcal{P})$ with respect to the basis $\ket{kk'}\bra{ll'}$ possess the following properties:
\begin{gather}
 \label{eqn::PropB}
(1) \ \mathcal{N}_{kk';ll'} = \mathcal{N}_{l'l;k'k}, \qquad 
(2) \ \mathcal{N}_{kk';ll'} = \mathcal{N}_{kl;k'l'}, \qquad
(3) \ \mathcal{N}^*_{kk';ll'} = \mathcal{N}_{k'k;l'l}.
\end{gather}
By counting the number of remaining independent entries in the matrix $\mathcal{N}$, one can deduce that the vector space of matrices with the properties laid out in Eq.~\eqref{eqn::PropB} is $\frac{1}{4}d_\mathcal{S}^2\left(d_\mathcal{S}+1\right)^2$-dimensional. In principle, it remains to be shown that $\text{Span}(\mathcal{P})$ actually coincides with this vector space (the matrix $\mathcal{N}$ given in Eq.~\eqref{eqn::vecProj} could in principle have further symmetries than the ones stated in Eq.~\eqref{eqn::PropB}). However, for the qubit case, a set of $\frac{1}{4}d_\mathcal{S}^2\left(d_\mathcal{S}+1\right)^2=9$ pure states $Q_i$ that yields linearly independent projective maps has been constructed in Ref.~\cite{kuah_how_2007}:
\begin{gather}
 \label{eqn::PureStates}
 Q_{(j,\pm)} = \frac{1}{2}(\mathbbm{1} \pm \sigma_j) \quad \text{and} \quad
 Q_{(k+l+1,+)} = \frac{1}{2}\left(\mathbbm{1} + \frac{1}{\sqrt{2}}\sigma_k + \frac{1}{\sqrt{2}}\sigma_l\right), 
\end{gather}
where $k<l$ and $\{\sigma_1,\sigma_2,\sigma_3\} \equiv \{\sigma_x,\sigma_y,\sigma_z\}$. For other low-dimensional cases, it is possible to numerically find $\frac{1}{4}d_\mathcal{S}^2\left(d_\mathcal{S}+1\right)^2$ linearly independent projective maps.  We leave the general statement regarding the dimension as a conjecture. The restricted process tensor for an $N$-step process can then be constructed by determining the output states for all $[\frac{1}{4}d_\mathcal{S}^2\left(d_\mathcal{S}+1\right)^2]^N$  possible combinations of basis projections performed at the time steps $\{t_k\}_{k=0}^{N-1}$.

\section{Detection of correlations \texorpdfstring{$\Rightarrow$}{} non-Markovian dynamics}
\label{App::ProofMark}
Here we prove that if correlations are detectable by means of local operations, the underlying process is non-Markovian. 

Let $\rho_{\mathcal{SE}}\left(\mathbf{A}_{N-2:0}\right) \equiv \rho_{\mathcal{SE}} = \rho_\mathcal{S} \otimes \rho_\mathcal{E} + \chi_{\mathcal{SE}}$ denote the system-environment density matrix at the time step $t_{N-1}$ and $\hat{\mathcal{U}}$ the total unitary dynamics from $t_{N-1}$ to the final step $t_N$. If the correlations present in $\rho_\mathcal{SE}$ are detectable by a local operation $\hat{\mathcal{A}}$, we have 
\begin{gather}
 \label{eqn::detect}
 \tr_{\mathcal{E}}\left\{\hat{\mathcal{U}}\left[(\hat{\mathcal{A}}\otimes \mathcal{I}_{\mathcal{E}})\left[\chi_{\mathcal{SE}}\right]\right]\right\} \neq \mathbf{0}.
\end{gather}
Note that this is true independent of the witness that was used to detect the correlations. Let $\left\{\Pi_\mu\right\}_{\mu=1}^{2d_\mathcal{S}^2 - d_\mathcal{S}}$ be a set of projectors on the pure states $\left\{\ket{k}\right\}$, $\left\{\frac{1}{\sqrt{2}}\left(\ket{k} + \ket{l}\right)\right\}$, $\left\{\frac{1}{\sqrt{2}}\left(\ket{k} + i\ket{l}\right)\right\}$,  $\allowbreak \left\{\frac{1}{\sqrt{2}}\left(\ket{k} - \ket{l}\right)\right\}$ and $\left\{\frac{1}{\sqrt{2}}\left(\ket{k} - i\ket{l}\right)\right\}$, where $k$ and $l$ run from $1$ to $d_\mathcal{S}$ and $k<l$ is implied, and let $\left\{P_m\right\}_{m=1}^{d_\mathcal{S}^2}$ be a set of density matrices that constitutes a basis of $\mathcal{B}\left(\mathcal{H}_\mathcal{S}\right\}$. The set $\left\{\Pi_\mu\right\}$ forms an overcomplete basis of $\mathcal{B}\left(\mathcal{H}_\mathcal{S}\right)$ with the appealing property that $\sum_{\mu=1}^{2d_\mathcal{S}^2 - d_\mathcal{S}} \Pi_\mu = \left(2d_\mathcal{S} - 1\right) \mathbbm{1}_\mathcal{S}$, where $\mathbbm{1}_\mathcal{S}$ is the $d_\mathcal{S} \times d_\mathcal{S}$ identity matrix.  Every local operation $\mathcal{A} = \sum_{m,\mu}\alpha_{m\mu} P_m \otimes \Pi_\mu^\text{T} $ can be written as a linear combination of causal breaks and hence 
\begin{gather}
 \label{eqn::detect2}
 \tr_{\mathcal{E}}\left\{\hat{\mathcal{U}}\left[\left(\mathcal{A}\otimes \mathcal{I}_{\mathcal{E}}\right)\left[\chi_{\mathcal{SE}}\right]\right]\right\}
 =\sum_{m,\mu} \alpha_{m\mu} \tr_{\mathcal{E}}\left\{\hat{\mathcal{U}}\left[P_m\otimes \tr_{\mathcal{S}} \left(\Pi_\mu \chi_{\mathcal{SE}}\right)\right]\right\} \equiv \sum_{m,\mu} \alpha_{m\mu} \sigma_{m\mu}  \neq \mathbf{0}.
\end{gather}
Consequently, there is at least one pair $\left(m_0,\mu_0\right)$, for which $\sigma_{m_0\mu_0} \neq \mathbf{0}$. 

In what follows, we set $P_{m_0}\equiv P$ and $\sigma_{m_0\mu} \equiv \sigma_\mu$ and show that there exist two different causal breaks $P\otimes \Pi_\nu^\text{T}$ and $P\otimes \Pi_\xi^\text{T}$, such that the final reduced state after these causal breaks are applied differ, \textit{i.e.},
\begin{gather}
 \tr_{\mathcal{E}}\{\hat{\mathcal{U}}[P\otimes \tr_\mathcal{S}(\Pi_\nu \rho_\mathcal{SE})]\} 
\not \propto \ 
 \tr_{\mathcal{E}}\{\hat{\mathcal{U}}[P\otimes \tr_\mathcal{S}(\Pi_\xi \rho_\mathcal{SE})]\}\, ,
 \label{eqn::CausalBr}
\end{gather}
which means that the process is non-Markovian. By setting 
\begin{gather}
p_\nu \eta \equiv \tr_{\mathcal{E}}\left\{\mathcal{U}\left[P\otimes \tr_\mathcal{S}\left(\Pi_\nu (\rho_{\mathcal{S}} \otimes \rho_{\mathcal{E}})\right)\right]\right\}, 
\end{gather}
where $p_\nu = \tr[\Pi_\nu\rho_\mathcal{S}]$ is the probability to measure the outcome corresponding to $\Pi_\nu$ given the total state $\rho_{\mathcal{S}} \otimes \rho_{\mathcal{E}}$, Eq.~\eqref{eqn::CausalBr} can be rewritten as 
\begin{gather}
 \label{eqn::CausalShort}
 p_\nu \eta + \sigma_\nu \not \propto p_\xi \eta + \sigma_\xi. 
\end{gather}
The process is only Markovian if the left and right sides of Eq.~\eqref{eqn::CausalShort} are proportional for all pairs $(\nu,\xi)$.
We note that $\tr\left(\eta\right) = 1$, while, due to the fact that $\tr_{\mathcal{E}}\left(\chi_{\mathcal{SE}}\right) = \tr_{\mathcal{S}}\left(\chi_{\mathcal{SE}}\right) = \mathbf{0}$, it is straightforward to show that for all projections $\Pi_\mu$ we have $\tr\left(\sigma_\mu\right)=0$. Consequently, proportionality in Eq.~\eqref{eqn::CausalShort} would only hold if $\sigma_\nu = p_\nu/p_\xi \cdot \sigma_\xi$. We choose $\Pi_\xi$ such that $p_\xi \neq 0$ (this is always possible). If $p_\nu=0$, the two sides of Eq.~\eqref{eqn::CausalShort} are not proportional and the process is non-Markovian. Otherwise, the ratio $p_\nu/p_\xi$ has to be positive. Given that there is at least one $\sigma_{\mu_0} \neq \mathbf{0}$, it is possible to find a projection $\Pi_\nu$ such that Eq.~\eqref{eqn::CausalShort} holds, \textit{i.e.}, it is possible to find a projection $\Pi_\nu$ such that $\sigma_\nu \not \propto \sigma_\xi$ or $\sigma_\nu = \lambda \sigma_\xi$ with $\lambda<0$. We assume the opposite and show that it leads to a contradiction. 

Let all $\sigma_\mu$ be proportional to each other, \textit{i.e.} $\sigma_\mu = \beta_\mu \sigma$ for all $\mu$, where $\sigma\neq \mathbf{0}$ is some traceless matrix and $\beta_\mu > 0$. If this holds, we obtain
\begin{align}
\notag
 \sum_{\mu=1}^{2d_\mathcal{S}^{2}-d_\mathcal{S}} \sigma_\mu &=  \sum_{\mu=1}^{2d_\mathcal{S}^{2}-d_\mathcal{S}} \beta_\mu \, \sigma = \sum_{\mu=1}^{2d_\mathcal{S}^{2}-d_\mathcal{S}}\tr_{\mathcal{E}}\{\hat{\mathcal{U}}[P\otimes \tr_{\mathcal{S}} (\Pi_\mu \chi_{\mathcal{SE}})]\} \\ &=\left(2d_\mathcal{S}-1\right)\tr_{\mathcal{E}}\left\{\hat{\mathcal{U}}\left[P\otimes \tr_{\mathcal{S}} \left(\chi_{\mathcal{SE}}\right)\right]\right\} = \mathbf{0},
\end{align}
which implies that $\sum_\mu^{2d_\mathcal{S}^{2}-d_\mathcal{S}} \beta_\mu = 0$. Therefore, given that at least one of the factors $\beta_\mu$ has to differ from zero, there exist indices $\mu_1,\mu_2$ such that $\beta_{\mu_1}>0$ and $\beta_{\mu_2}<0$ and $\sigma_{\mu_1} = \beta_{\mu_1}/\beta_{\mu_2} \cdot \sigma_{\mu_2} \equiv \lambda \sigma_{\mu_2} $ with $\lambda<0$, which concludes the proof.\\

\twocolumngrid
\bibliographystyle{apsrev4-1}
\bibliography{references}
\end{document}